\documentclass[prd,nofootinbib,superscriptaddress,preprint]{revtex4}
\usepackage[T1]{fontenc}
\usepackage{amsmath,amssymb}
\usepackage{epsfig}
\usepackage{graphicx}
\usepackage[usenames,dvipsnames]{color}
\usepackage{slashed}
\usepackage{bbold}
\usepackage[colorlinks,citecolor=blue]{hyperref}
\usepackage{pdfpages}
\usepackage{color}
\usepackage{subfig}

\newcommand{\R}{\mbox{\tiny$R$}}

\begin{document}

\title{Planck Scale Origin of Non-zero $\theta_{13}$ and Super-WIMP Dark Matter} 

\author{Debasish Borah}
\email{dborah@iitg.ac.in}
\affiliation{Department of Physics, Indian Institute of Technology
Guwahati, Assam 781039, India}
\author{Biswajit Karmakar}
\email{biswajit@prl.res.in}
\affiliation{Theoretical Physics Division, Physical Research Laboratory, Ahmedabad 380009, India}
\author{Dibyendu Nanda}
\email{dibyendu.nanda@iitg.ac.in}
\affiliation{Department of Physics, Indian Institute of Technology
Guwahati, Assam 781039, India}

\begin{abstract}
We study a discrete flavour symmetric scenario for neutrino mass and dark matter under the circumstances where such global discrete symmetries can be explicitly broken at Planck scale, possibly by gravitational effects. Such explicit breaking of discrete symmetries mimic as Planck suppressed operators in the model which can have non-trivial consequences for neutrino and dark matter sectors. In particular, we study a flavour symmetric model which, at renormalisable level, gives rise to tri-bimaximal type neutrino mixing with vanishing reactor mixing angle $\theta_{13}=0$, a stable inert scalar doublet behaving like a weakly interacting massive particle (WIMP) and a stable singlet inert fermion which does not interact with any other particles. The introduction of Planck suppressed operators which explicitly break the discrete symmetries, can give rise to the generation of non-zero $\theta_{13}$ in agreement with neutrino data and also open up decay channels of inert scalar doublet into singlet neutral inert fermion leading to the realisation of the super-WIMP dark matter scenario. We show that the correct neutrino phenomenology can be obtained in this model while discussing three distinct realisation of the super-WIMP dark matter scenario.
\end{abstract}

\maketitle 

\section{Introduction}
Non-zero but tiny neutrino mass and large leptonic mixing have become a well established fact by now, thanks to several experimental efforts in the last two decades \cite{Fukuda:2001nk, Ahmad:2002jz, Ahmad:2002ka, Abe:2008aa, Abe:2011sj, Abe:2011fz, An:2012eh, Ahn:2012nd, Adamson:2013ue}. For a review of neutrino mass and mixing, please see \cite{Tanabashi:2018oca}. Among these experiments, the relatively new ones namely, the T2K \cite{Abe:2011sj}, the Double Chooz \cite{Abe:2011fz}, the Daya Bay \cite{An:2012eh}, the RENO \cite{Ahn:2012nd} and the MINOS \cite{Adamson:2013ue} have not only reconfirmed the fact that neutrinos oscillate by making the measurements of mass squared differences and mixing angles more precise, but also discovered a non-zero value of of reactor mixing angle $\theta_{13}$ which was thought to be almost vanishing earlier. For a recent global fit of neutrino oscillation data, we refer to \cite{deSalas:2017kay, Esteban:2018azc}. Apart from neutrino oscillation experiments, cosmology experiment like Planck also constrains the neutrino sector by putting an upper bound on the sum of absolute neutrino masses $\sum_i \lvert m_i \rvert < 0.11$ eV \cite{Aghanim:2018eyx}. As indicated by these global fits, in spite of these experimental developments, there still remains several unknowns in the neutrino sector like the nature of neutrinos: Dirac or Majorana, mass hierarchy: normal or inverted, CP violating phases etc. Apart from these, the origin of light neutrino mass also remains unknown in the standard model (SM) as the Higgs field do not have any renormalisable coupling to the neutrinos due to the absence of right handed neutrinos. If we go beyond renormalisable level, then it is possible to generate light neutrino mass of Majorana type via a dimension five Weinberg operator \cite{Weinberg:1979sa} of type $(L L H H)/\Lambda$ where $\Lambda$ is an unknown cutoff scale somewhere above the electroweak scale. Usual seesaw models \cite{Minkowski:1977sc, GellMann:1980vs, Mohapatra:1979ia, Schechter:1980gr} for light neutrino mass attempt to provide a dynamical origin of the Weinberg operator by introducing new fields like heavy right handed neutrinos. Apart from the sub-eV mass, well below the electroweak scale, another puzzling feature related to neutrinos is their large mixing, in sharp contrast with small mixing angles in the quark sector. This has also motivated the study of different flavour symmetry models that can predict such large mixing patterns. One of the very popular flavour symmetric scenarios is the one that predicts a $\mu-\tau$ symmetric light neutrino mass matrix. However, such a scheme which predicts $\theta_{13} = 0, \theta_{23} = \frac{\pi}{4}$ and different values of $\theta_{12}$ depending upon the particular realisation of this symmetry \cite{Xing:2015fdg}, has already been ruled out by recent neutrino experiments. Among such realisations, the tri-bimaximal (TBM) mixing \cite{Harrison:2002kp, Xing:2002sw, Harrison:2002et, Harrison:2003aw, Harrison:2004he} which predicts $\theta_{12}=35.3^o$ has been the most popular one. This mixing, which was consistent with neutrino data before the discovery of non-zero $\theta_{13}$ can be realised naturally within several flavour symmetric models based on non-Abelian discrete groups \cite{Ishimori:2010au, Grimus:2011fk, King:2013eh, Altarelli:2010gt}. Among such flavour symmetric models, the discrete group $A_4$, group of even permutations of four objects, can reproduce the TBM mixing in the most economical way \cite{Ma:2001dn, Babu:2002dz, Hirsch:2003dr, Ma:2004zv, Ma:2004zd, Chen:2005jm, Ma:2005sha, Zee:2005ut, Ma:2005mw, Ma:2005qf, Altarelli:2005yx}. However, in order to be consistent with the present neutrino data, such TBM or $\mu-\tau$ symmetric scenarios have to be corrected to generate non-zero value of $\theta_{13}$. There have been several attempts in that direction, some of which can be found in \cite{Kang:2005bg, Shimizu:2011xg, King:2011zj, Antusch:2011ic, King:2011ab, Gupta:2011ct, Ge:2011qn, Ge:2010js, Ge:2011ih, Liao:2012xm, Xing:2010pn, Adhikary:2006wi, Ma:2011yi, Altarelli:2012bn,  Karmakar:2014dva, Chen:2012st, Borah:2013jia, Borah:2013lva, Borah:2014fga, Borah:2014bda, Kalita:2014mga, Mukherjee:2017pzq, Borah:2017qdu} and references therein.

Apart from non-zero neutrino mass and large leptonic mixing, another observed phenomena that has propelled serious hunt for beyond standard model (BSM) physics is the presence of non-baryonic form of matter, or the so called dark matter (DM) in large amount in the present universe. Apart from the longstanding astrophysical evidences \cite{Zwicky:1933gu, Rubin:1970zza, Clowe:2006eq}, the recent cosmology experiment Planck suggests that almost $26\%$ of the present 
Universe's energy density is in the form of DM while only around $5\%$ is the 
usual baryonic matter leading the rest of the energy budget to mysterious dark 
energy \cite{Aghanim:2018eyx}. In terms of density 
parameter $\Omega_{\rm DM}$ and $h = \text{Hubble Parameter}/(100 \;\text{km} ~\text{s}^{-1} 
\text{Mpc}^{-1})$, the present DM abundance is conventionally reported as \cite{Aghanim:2018eyx}:
$\Omega_{\text{DM}} h^2 = 0.120\pm 0.001$
at 68\% CL. Since none of the SM particles can satisfy the requirements for being a DM candidate, several BSM proposals have been put forward out of which the weakly interacting massive particle (WIMP) paradigm is perhaps the most popular or the most widely studied one. The coincidence that a stable or cosmologically long lived particle having electroweak scale mass and interactions fulfilling the criteria for observed DM abundance is often referred to as the \textit{WIMP Miracle}. However, if such particles having electroweak scale mass and interactions really exist in the universe with such a large density, they are expected to pass through the detectors of several DM direct detection experiments giving rise to nuclear recoils. However, there have been no detection of particle DM at any experiments. 
The direct detection experiments like LUX \cite{Akerib:2016vxi}, PandaX-II \cite{Tan:2016zwf, Cui:2017nnn} and Xenon1T \cite{Aprile:2017iyp, Aprile:2018dbl} have continued to produce null results so far. Similar null results follow from collider searches at the large hadron collider (LHC) \cite{Kahlhoefer:2017dnp} as well as the indirect detection frontiers \cite{Ahnen:2016qkx}. The typical indirect detection experiments excess of antimatter, gamma rays or neutrinos, originating perhaps 
from dark matter annihilations (for stable DM) or decay (for long lived DM) and no convincing signal has been observed at any experiment operating in this frontier. Such null results from WIMP searches have led to the proposals of several alternative frameworks of DM, specially to scenarios where the interaction between DM and visible sector could be much more weaker than what it is in the WIMP paradigm. In fact, if the coupling of DM to visible sector is sufficiently weak, then DM can never be produced thermally in the universe requiring a non-thermal origin for its relic \cite{Hall:2009bx}. In such scenarios, the DM has negligible initial number density in the early universe and is produced from out of equilibrium decay or scattering of visible sector particles. The production mechanism for non-thermal DM
is known as freeze-in and the candidates of non-thermal DM produced
via freeze-in are often classified into a group called
Freeze-in (Feebly interacting) massive particle (FIMP). For a recent review of  
this DM paradigm, please see \cite{Bernal:2017kxu}. The tiny couplings between DM and visible sector can be naturally realised either by higher dimensional operators \cite{Hall:2009bx, Elahi:2014fsa, McDonald:2015ljz} or through some UV complete renormalisable theories \cite{Biswas:2018aib}. While typical FIMP does not have much direct detection prospects and WIMP direct detection has so far failed, there exists a scenario which is a combination of both and have interesting detection prospects in terms of secondary particles. This is known as super-WIMP scenario \cite{Feng:2003uy} where a metastable WIMP decays into a super-weakly interacting dark matter at late epochs. Although DM still has feeble interactions with the visible sector, the metastable WIMP has sizeable interactions and can be detected as, for example, long lived BSM particle at collider experiments \cite{Lee:2018pag}. This scenario was also adopted in the context of neutrino mass models in several works including \cite{Borah:2017dfn, Borah:2018gjk, Borah:2019bdi}. In most of such models there exists some exact or approximate discrete symmetries to stabilise the DM or to give rise to a long lived metastable WIMP. 

In this work, we try to find a common thread linking the discrete symmetries in neutrino sector and DM sector. We consider the neutrino sector to have an exact $\mu-\tau$ symmetry while the dark sector has an exact $Z_2$ symmetry. We find a common source of breaking of these discrete symmetries through Planck scale suppressed operators by using the well known argument that any generic theories of quantum gravity should not respect global symmetries: both discrete and continuous \cite{Abbott:1989jw, Kallosh:1995hi, Hawking:1974sw}. Recently, this argument was used to realise light neutrino mass from Planck scale lepton number breaking \cite{Ibarra:2018dib}. Effects of such Planck scale breaking of discrete symmetries on light neutrino parameters was studied within the context of left right symmetric model a few years back in \cite{Borah:2013mqa}. Here we study the consequences of such Planck scale breaking of discrete symmetries on neutrino sector which generates non-zero $\theta_{13}$, as required by present neutrino data. At the same time, such breaking also makes it possible to realise the super-WIMP scenario by assisting the decay of a metastable WIMP to a super-weakly interacting dark matter sector. We consider the presence of discrete flavour symmetries at a scale between electroweak and Planck scale to dictate the patterns of neutrino mixing, neutrino mass as well as DM dynamics while the Planck scale suppressed terms break these global discrete symmetries affecting the neutrino as well as DM sector. Presence of an intermediate scale makes the Planck scale suppressed terms relevant in the discussion which otherwise would have been of marginal importance through the usual Weinberg operator 
$(L L H H)/M_{\rm Pl}$ that remains several order of magnitudes below typical neutrino mass scale, in the absence of any intermediate scale between electroweak scale and $M_{\rm Pl}$. We show that the scenario can have interesting predictions in both neutrino and DM sectors. In the DM sector, we also discuss the corresponding phenomenology in the presence of a spontaneously broken gauged sector which forbids several Planck suppressed terms present in the scenario where DM sector is charged only under some discrete global symmetries. Our framework can be generalised to a Planck scale origin of lepton number violation and hence neutrino mass, similar to \cite{Ibarra:2018dib} along with super-WIMP dark matter. We however, restrict ourselves to discussing the origin of non-zero $\theta_{13}$ in this work, without assuming a global lepton number symmetry at renormalisable level of the theory.

This paper is organised as follows. In section \ref{sec:model} we discuss our flavour symmetric model and possible Planck suppressed operators. In section \ref{sec2} we discuss neutrino mass and mixing followed by super-WIMP dark matter phenomenology in section \ref{sec3}. We finally conclude in section \ref{sec4}.

\section{$A_4$ Model with TBM mixing}
\label{sec:model}
In this section we outline the flavour model based on non-Abelian discrete group $A_4$ augmented by a $Z_4$ symmetry for dark sector. A brief summary of $A_4$ group, its representations and product rules are given in appendix \ref{appen1}. The particle content of the model, relevant for discussing the lepton sector and dark matter sector is shown in table \ref{tab:a}. Apart from the SM lepton doublets and charged lepton singlets, there are three right handed neutrinos required for implementing the type I seesaw mechanism \cite{Minkowski:1977sc, GellMann:1980vs, Mohapatra:1979ia, Schechter:1980gr}. In addition to the usual SM Higgs doublet, there exist three additional Higgs doublets required to generate Dirac mass term for neutrinos. There also exists two more scalar fields responsible for generating the desired right handed neutrino mass matrix and also to break the $A_4$ flavour symmetry spontaneously. Two more fields one scalar and one fermion, both charged under the $Z_4$ symmetry are included in order to achieve the desired DM phenomenology.

\begin{table}[h]
\centering
\begin{tabular}{|c|c|c|ccccc|cc|}
\hline
 Fields & $L_e, L_{\mu}, L_{\tau}$  & $e_{\R}, \mu_{\R}, \tau_{\R}$ &  $H$ & $H_T$ & $\phi_N$&$\xi$ & $N_{R}$& $\psi$& $\eta$  \\
\hline
$A_{4}$ & 1, $1^{\prime}, 1^{\prime \prime}$ & 1,$1''$,$1'$ & 1 & 3 & 3 &1& 3 &1 & 1 \\
\hline
$Z_{4}$ & 1 & 1 & 1 & 1 &1 & 1&1 & i & -1 \\
\hline
$SU(2)_L$ & 2 & 1 & 2 & 2 &1 &1& 1 & 1 & 2 \\
\hline
$U(1)_Y$ & $\frac{1}{2}$ & 1 & $-\frac{1}{2}$ & $-\frac{1}{2}$ &0 &0& 0 & 0 & $-\frac{1}{2}$ \\
\hline
\end{tabular}
\caption{\label{tab:a} Field content and transformation properties under
$A_4 \times Z_4 $ symmetry of the model. }
\end{table}
The renormalisable Yukawa Lagrangian for leptons can be written as 
\begin{align}
\mathcal{L}_Y & \supset Y_e \overline{L_e} H e_R +Y_{\mu} \overline{L_{\mu}} H \mu_R + Y_{\tau} \overline{L_{\tau}} H \tau_R + (Y_{\nu 1} \overline{L_e}+Y_{\nu 2} \overline{L_{\mu}}+Y_{\nu 3} \overline{L_{\tau}}) \tilde{H}_T N_R \nonumber \\
& + (Y_{\xi}\xi+Y_N \phi_N) N_R N_R +{\rm h.c.}
\end{align}
where $\tilde{H}_T = \tau_2 H^*_T$. Due to the additional $Z_4$ symmetry, the fields $\psi, \eta$ do not have any renormalisable coupling with SM leptons or right handed neutrinos. We assume $\psi$ be vector like so that a bare mass term $M_{\psi} \bar{\psi} \psi$ is allowed in the Lagrangian. As mentioned earlier, gravity is not supposed to obey such global symmetries and hence we can write down Planck suppressed terms which explicitly break both $A_4$ and $Z_4$. The dimension five terms involving one of the $A_4$ triplet flavon $\phi_N$ which explicitly break the discrete symmetries but preserve the gauge symmetries of the standard model can be written as
\begin{align}
\mathcal{L}_{\rm Planck} & \supset \frac{1}{M_{\rm Pl}} \bigg [ \sum _{i, \alpha, \beta} \phi_{Ni} (Y^{'}_1 \overline{L_{\alpha}} H e_{\beta} + Y^{'}_2 \overline{L_{\alpha}} \eta e_{\beta}+Y^{''}_1 \overline{L_{\alpha}} \tilde{H} \psi + Y^{''}_2 \overline{L_{\alpha}} \tilde{\eta} \psi) \nonumber \\
& +\sum_{i, \alpha, \beta} \phi_{Ni}(Y'_3 \overline{L_{\alpha}} \tilde{H} N_{\beta} +Y'_4 \overline{L_{\alpha}} \tilde{\eta} N_{\beta}) + \tilde{Y_1}H \eta^{\dagger} \psi \psi \bigg ] +{\rm h.c.}
\label{plH}
\end{align}

We can also write down the dimension five terms simultaneously involving two of the $A_4$ triplet flavons $H_T, \phi_N$ which explicitly break the discrete symmetries but preserve the gauge symmetries of the standard model. They can be written as
\begin{align}
\mathcal{L}_{\rm Planck} & \supset \frac{1}{M_{\rm Pl}} \bigg [ \sum _{i, j, \alpha, \beta} \phi_{Ni} (\tilde{Y^{'}}_1 \overline{L_{\alpha}} H_{Tj} e_{\beta} + \tilde{Y^{''}}_1 \overline{L_{\alpha}} \tilde{H}_{Tj} \psi +\tilde{Y'}_3 \overline{L_{\alpha}} \tilde{H}_{Tj} N_{\beta}) +\tilde{Y_2} H_{Tj} \eta^{\dagger} \psi \psi \bigg ] +{\rm h.c.}
\label{plHT}
\end{align}

One can prevent the coupling of $\eta, \psi$ with standard model leptons at dimension five level by considering an additional $U(1)_X$ gauge symmetry under which $\psi$ is a Dirac fermion with charge $n_1$ while $\eta$ has a charge $2n_1$. Thus, $\eta$ can behave like a next to lightest particle which decays only to $\psi$ due to the Planck scale effects on $Z_2$ symmetry breaking.

As discussed below, we choose a hierarchy of flavon vacuum expectation value (VEV)'s in order to achieve the desired phenomenology. While the $SU(2)_L$ singlet flavon VEV's can be arbitrary, the same does not apply to that of the triplet flavon which is also a doublet under $SU(2)_L$. In order to be in agreement with precesion electroweak measurements \cite{Tanabashi:2018oca}, the vev of the neutral components of $A_4$ singlet Higgs doublet $H (v_1)$ and $A_4$ triplet Higgs doublet $H_T$ namely, $v_i, i=2,3,4$ must satisfy $\sqrt{v^2_1+v^2_2+v^2_3+v^2_4} \approx 174$ GeV. The scalar potential involving $H, H_T$ only can be written as
\begin{eqnarray}
V(H,H_T) &=& -\mu_H^2 H^{\dagger}H 
 - \mu_{2}^2H_T^{\dagger} H_T +\lambda_1(H^{\dagger}H)^2 + \lambda_2 [H_T^{\dagger} H]_{1}^2+ \lambda_3 [H_T^{\dagger} H_T]_{1'}[H_T^{\dagger} H_T]_{1''} \nonumber \\
&&
 + \frac{\lambda_4}{2} ([H_T^{\dagger} H_T^{\dagger}]_{1'}[H_T H_T]_{1''}+[H_T^{\dagger} H_T^{\dagger}]_{1''}[H_T H_T]_{1'} ) + \lambda_5 [H_T^{\dagger} H_T^{\dagger}]_{1}[H_T H_T]_{1} \nonumber \\
 &&
 + \frac{\lambda_6}{2} ([H^{\dagger}_T H_T]_{3_1}[H_T^{\dagger} H_T]_{3_1}+{\rm h.c.})+ \lambda_7[H_T^{\dagger} H_T]_{3_1}[H_T^{\dagger} H_T]_{3_2}+ 
\lambda_8 [H_T^{\dagger} H_T^{\dagger}]_{3_1}[H_T H_T]_{3_2}\nonumber \\
&&
+
\lambda_{9} [H_T^{\dagger} H_T]_{1}H^{\dagger}H
+\lambda_{10}[H_T^{\dagger} H]_{3_1}[H^{\dagger} H_T]_{3_1} +\frac{\lambda_{11}}{2}([H_T^{\dagger} H_T^{\dagger}]_{1}HH +{\rm h.c.})\nonumber \\
&&
+
 \frac{\lambda_{12}}{2}([H_T^{\dagger} H_T]_{3_1}H_T^{\dagger}H +{\rm h.c.})
+\frac{\lambda_{13}}{2}([H_T^{\dagger} H_T]_{3_2}[H_T^{\dagger}H]_{3_1} +{\rm h.c.})\nonumber \\
&&
+\frac{\lambda_{14}}{2}([H_T^{\dagger} H_T^{\dagger}]_{3_1}H_T H +{\rm h.c.})+
\frac{\lambda_{15}}{2}([H_T^{\dagger} H_T^{\dagger}]_{3_2} H_T H +{\rm h.c.})
 \end{eqnarray}
which is similar to the scalar potential of a multi Higgs doublet model with additional complications due to the non-trivial $A_4$ transformations of $A_4$. Similarly, one can write the scalar potential involving $SU(2)_L$ singlet scalar fields as well as their interactions with $H, H_T$. Although a complete analysis of vacuum alignment \footnote{For details of vacuum alignment in supersymmetric and non-supersymmetric discrete flavour symmetric models, one may refer to \cite{Altarelli:2005yx, Feruglio:2009iu} and \cite{Holthausen:2011vd} respectively.} is beyond the scope of this work, it is possible to keep the VEV's of $H, H_T$ around the electroweak scale by suitable choices of their bare mass squared terms and quartic couplings while keeping their interaction with the singlet flavons (which acquire high scale VEV's) sufficiently small. The phenomenological and detection prospects of the components of $H_T$ will be similar to two or multi Higgs doublet models \cite{Branco:2011iw, Akeroyd:2016ymd} as the physical masses of all the components can be kept around the TeV scale.
\section{Lepton masses and mixings}
\label{sec2}

Up to leading order, owing to the $A_4$ symmetry, one can write the charged lepton mass matrix in the diagonal form as
\begin{equation}
M_{\ell}=\left(\begin{array}{ccc}
Y_e& 0&0\\
0& Y_{\mu} & 0 \\
0 & 0 & Y_{\tau}
\end{array}\right)v_1
\label{mclepton}
\end{equation}
where $v_1$ is the VEV of the neutral component of the Higgs doublet $H$. We use the $A_4$ product rules given in appendix \ref{appen1} in order to evaluate the forms of different mass matrices at renormalisable level. Considering the vacuum alignment of $A_4$ triplet scalar field
$H_T = (v_2, 0, 0)$, the Dirac neutrino mass matrix at leading order can be written as,  
\begin{equation}
M_D^0=\left(\begin{array}{ccc}
Y_{\nu 1}v_2& 0&0\\
0& 0 &Y_{\nu 2}v_2 \\
0 &Y_{\nu 3}v_2& 0
\end{array}\right), 
\label{mdirac}
\end{equation}
where, without any loss of generality, one can consider $Y_{\nu 1}=Y_{\nu 2}=Y_{\nu 3}$.

Considering triplet flavon vacuum alignment as $\langle \phi_N \rangle = u(1,1,1)$ and $\langle \xi \rangle = u_{\xi}$ the right
handed neutrino mass matrix at leading order is given by
\begin{equation}
M_{R}= \left(\begin{array}{ccc}
a+\frac{2 b}{3} &-\frac{b}{3} &-\frac{b}{3} \\
-\frac{b}{3} &\frac{2b}{3}  & a-\frac{b}{3}  \\
-\frac{b}{3} &a-\frac{b}{3} &\frac{2b}{3}
\end{array}\right),
\label{mright}
\end{equation}
where $a=2Y_{\xi}u_{\xi}$ and $b=2Y_{N}u$. Hence the light neutrino mass matrix, using the type I seesaw formula is
\begin{equation}
-M_{\nu}= M^0_D M^{-1}_{R} {M^0}_D^T= v^2_2 Y^2_{\nu 1} \left(\begin{array}{ccc}
\frac{3a+b}{3(a+b)} & \frac{b}{3(a+b)} & \frac{b}{3(a+b)} \\
\frac{b}{3(a+b)} & \frac{-b(2a+b)}{3(a+b)(a-b)} & \frac{3a^2+ab-b^2}{3(a+b)(a-b)} \\
\frac{b}{3(a+b)} & \frac{3a^2+ab-b^2}{3(a+b)(a-b)} & \frac{-b(2a+b}{3(a+b)(a-b)}
\end{array}\right). 
\label{mlight1}
\end{equation}
This $\mu-\tau$ symmetric light neutrino mass matrix can be diagonalised by the usual tri-bimaximal mixing matrix given by~\cite{Harrison:2002er}
\begin{equation}
U_{\rm TBM}=\left(\begin{array}{ccc}
\sqrt{\frac{2}{3}} & \frac{1}{\sqrt{3}} & 0 \\
\frac{-1}{\sqrt{6}} & \frac{1}{\sqrt{3}} &\frac{-1}{\sqrt{2}} \\
\frac{-1}{\sqrt{6}} & \frac{1}{\sqrt{3}} &\frac{1}{\sqrt{2}}
\end{array}\right),
\label{utb}
\end{equation}
predicting $\theta_{13}=0$ and maximal value of $\theta_{23}$. Present experimental observation, however, rules out $\theta_{13}=0$ scenario and has a preference towards higher octant for $\theta_{23}$ 
 ($i.e. \theta_{23}>45^{\circ}$)~\cite{Esteban:2018azc}. We therefore consider the Planck scale suppressed corrections to such scenarios so that the correct light neutrino phenomenology can be obtained along with the implications for dark sector.

The contribution to the light neutrino mass matrix and neutrino mixing from the Planck suppressed terms which break the discrete symmetries explicitly can arise in a variety of ways. For example,
\begin{itemize}
\item there can be contributions to light neutrino mass matrix of the type $(L L H H)/M_{\rm Pl}, (L L H_T H_T)/M_{\rm Pl}, (L L H_T H)/M_{\rm Pl}$, all of which remain at least $10^{-5}$ times smaller than the typical neutrino mass scale, in the absence of any intermediate scale between electroweak scale and $M_{\rm Pl}$.
\item there can be new contributions to charged lepton mass matrix of the type $\phi_{Ni} (Y^{'}_1 \overline{L_{\alpha}} H e_{\beta})/M_{\rm Pl} $ which can be suppressed by an additional factor $u/M_{\rm Pl}$ compared to the leading order contributions to charged lepton masses. But such corrections can, in principle, lead to deviations from diagonal charged lepton mass matrix. Usually, the leptonic mixing matrix is given in terms of the charged lepton diagonalising matrix $(U_l)$ and light neutrino diagonalising matrix $U_{\nu}$ as $U = U^{\dagger}_l U_{\nu}$. In the simple case where the charged lepton mass matrix is diagonal which is true in our model at tree level, we can have $U_l = \mathbb{1}$. Therefore we can write $U = U_{\nu}$. But $U_l$ may become non-trivial after the corrections are added and it will mimic as a correction in the leading order leptonic mixing which is TBM type.
\item there can be corrections to the Dirac neutrino mass matrix of the type $\phi_{Ni}(Y'_3 \overline{L_{\alpha}} \tilde{H} N_{\beta}) /M_{\rm Pl}$ which can again be suppressed by an additional factor $u/M_{\rm Pl}$ compared to the leading order contributions. Such corrections will propagate to the light neutrino mass matrix through the type I seesaw formula.
\item there can be corrections to the right handed neutrino mass matrix of the type $ (Y'_{\xi}\xi^2+Y'_N \phi^2_N) N_R N_R/M_{\rm Pl}$ which can change the structure of $M_R$ from the $\mu-\tau$ symmetric leading order form mentioned before. Once again this correction will propagate to the light neutrino mass matrix via type I seesaw formula.
\end{itemize}
All these corrections can be simultaneously or separately sufficient to generate correct neutrino mixing. For representative purpose, we will consider corrections to the Dirac neutrino mass matrix only and that too via Higgs doublet $H$. Addition of other corrections will make the calculations complicated without any new insights. A more general analysis is beyond the scope of this work and left for future studies.

Now, following equation (\ref{plH}), the correction to the Dirac neutrino mass matrix now can be written as, 
\begin{equation}
M^1_D=\frac{u v_1}{M_{\rm Pl}}\left(\begin{array}{ccc}
(Y'_{3})_{11} & (Y'_{3})_{12} & (Y'_{3})_{13} \\
(Y'_{3})_{21} & (Y'_{3})_{22} &(Y'_{3})_{23} \\
(Y'_{3})_{31} &(Y'_{3})_{32} & (Y'_{3})_{33}
\end{array}\right), 
\label{mdirac1}
\end{equation}
In the present construction, following $A_4$ multiplication rules given in the appendix, we find that 
$(Y'_{3})_{11}=(Y'_{3})_{23}=(Y'_{3})_{32}$ , $(Y'_{3})_{12}=(Y'_{3})_{21}=(Y'_{3})_{33}$  and 
$(Y'_{3})_{13}=(Y'_{3})_{22}= (Y'_{3})_{31}$ . With such correction the effective Dirac neutrino mass matrix
can be written as (assuming $Y_{\nu 1}=Y_{\nu 2}=Y_{\nu 3}$)
\begin{eqnarray}
M_D=M^0_{D}+M_{D}^1&=&\left(\begin{array}{ccc}
Y_{\nu 1}v_2& 0&0\\
0& 0 &Y_{\nu 1}v_2 \\
0 &Y_{\nu 1}v_2& 0
\end{array}\right)+\frac{u v_1}{M_{\rm Pl}}\left(\begin{array}{ccc}
(Y'_{3})_{11} & (Y'_{3})_{12} & (Y'_{3})_{13} \\
(Y'_{3})_{12} & (Y'_{3})_{13} &(Y'_{3})_{11} \\
(Y'_{3})_{13} &(Y'_{3})_{11} & (Y'_{3})_{12}
\end{array}\right),\nonumber\\
&=&v\left(\begin{array}{ccc}
y_1& y_2&y_3\\
y_2& y_3 &y_1 \\
y_3 &y_1& y_2
\end{array}\right), 
\label{mdirac1}
\end{eqnarray}
where $y_1=Y_{\nu 1}+\frac{u }{M_{\rm Pl}}(Y'_{3})_{11}$, $y_2=\frac{u }{M_{\rm Pl}}(Y'_{3})_{12}$ 
and $y_3=\frac{u }{M_{\rm Pl}}(Y'_{3})_{13}$ with $v=v_2=v_1$.

Now, along with the Planck suppressed operator contribution in Dirac neutrino mass, the light neutrino mass originating from type I seesaw can be written as 
\begin{equation}
-M_{\nu}= M_D M^{-1}_{R} M_D^T. 
\label{mlight2}
\end{equation}
After a rotation by $U_{\rm TBM}$, the light neutrino mass matrix can be written as 
\begin{eqnarray}
M'_{\nu}&=& U_{\rm TBM}^T M_{\nu} U_{\rm TBM}\nonumber\\
&=&\left(\begin{array}{ccc}
\frac{-v^2(ax_2+2bx_3)}{a^2-b^2}& 0&\frac{-\sqrt{3}v^2x_1}{a^2-b^2}\\
0& \frac{v^2x_4}{a} &0 \\
\frac{-\sqrt{3}v^2x_1}{a^2-b^2} &0& \frac{v^2(ax_2-2bx_3)}{a^2-b^2}
\end{array}\right). 
\label{mTB}
\end{eqnarray}
where 
\begin{eqnarray}
 x_1&=&(2y_1-y_2-y_3)(y_2-y_3)\\
 x_2&=&-2y_1^2+y_2^2-4y_2y_3+y_3^2+2y_1(y_2+y_3)\\
 x_3&=&y_1^2+y_2^2-y_2y_3+y_3^2-y_1(y_2+y_3)\\
 x_4&=&(y_1+y_2+y_3)^2. 
\end{eqnarray}
Clearly an additional rotation in the 13 plane is required to diagonalise the above matrix via the relation 
\begin{equation}
 U_{13}^{T}{M'_{\nu}}U_{13}={\rm diag} 
(m_1e^{i\gamma_1},m_2e^{i\gamma_2},m_3e^{i\gamma_3})
\end{equation}
with
\begin{eqnarray}\label{u13}
U_{13}=\left(
\begin{array}{ccc}
 \cos\theta               & 0 & \sin\theta{e^{-i\psi}} \\
     0                    & 1 &            0 \\
 -\sin\theta{e^{i\psi}} & 0 &        \cos\theta
\end{array}
\right)  
\end{eqnarray}
where $m_1, m_2, m_3$ are the real positive neutrino mass eigenvalues and $\gamma_{1,2,3}$ are the respective Majorana phases. Therefore the effective neutrino mixing matrix (with diagonal charged 
lepton sector), can be written as 
\begin{equation}\label{unu}
 U_{\nu}=U_{\rm TBM}U_{13}U_{m}
\end{equation}
where $U_m$ is the diagonal Majorana phase matrix given by $U_m={\rm diag} (e^{i\gamma_1},e^{i\gamma_2},e^{i\gamma_3})$.
Comparing this with the Pontecorvo-Maki-Nakagawa-Sakata (PMNS) mixing matrix 
\begin{equation}
U_{\text{PMNS}}=\left(\begin{array}{ccc}
c_{12}c_{13}& s_{12}c_{13}& s_{13}e^{-i\delta}\\
-s_{12}c_{23}-c_{12}s_{23}s_{13}e^{i\delta}& c_{12}c_{23}-s_{12}s_{23}s_{13}e^{i\delta} & s_{23}c_{13} \\
s_{12}s_{23}-c_{12}c_{23}s_{13}e^{i\delta} & -c_{12}s_{23}-s_{12}c_{23}s_{13}e^{i\delta}& c_{23}c_{13}
\end{array}\right)P,
\label{PMNS}
\end{equation}
one can obtain the correlation between the model parameters and neutrino oscillation parameters. In the above PMNS mixing matrix, $P (={\rm diag} (1,e^{i\alpha_{21}/2},e^{i\alpha_{31}/2}))$ is the Majorana phase matrix where $\alpha_{21}=(\gamma_1-\gamma_2)$ and $\alpha_{31}=(\gamma_1-\gamma_3)$ are the physical Majorana phases after rotating away one common phase which is irrelevant. 
The rotation parameter $\theta$ in the rotation matrix \eqref{u13} can be obtained as
\begin{eqnarray}
 \tan 2\theta=\pm \sqrt{3}\frac{x_1}{x_2}, 
\end{eqnarray}
with $\psi=0,\pi$ respectively yielding maximal value for the Dirac CP phase $\delta$. 
The construction of the present set-up is such that we only have considered the Planck scale suppressed terms only for the Dirac 
neutrino mass matrix. Once we allow correction terms for the charged lepton and/or right handed neutrino mass matrix, non-maximal 
values for the Dirac CP phase $\delta$ becomes allowed~\cite{Karmakar:2014dva, Karmakar:2015jza, Karmakar:2016cvb}. Comparing the neutrino mixing matrix of our scenario given in (\ref{unu}) with the standard parametric form of mixing matrix given in (\ref{PMNS}), the neutrino mixing 
angles can be evaluated as~\cite{Karmakar:2014dva} 
\begin{eqnarray}
 \sin^2\theta_{13}&=&\frac{2}{3}|\sin\theta|^2\\
 \sin^2\theta_{12}&=&\frac{1}{3(1- \sin^2\theta_{13})}\\
 \sin^2\theta_{23}&=&\frac{1}{2}+\frac{1}{\sqrt{2}}\sin\theta_{13}\sin\psi. 
\end{eqnarray}
Thus, the neutrino mixing angles are function of the angle $\theta$ which depends upon $x_{1,2}$ and they further depend upon
the coupling constants $y_{1,2,3}$ as described before. As mentioned earlier, in the present set-up we have considered the 
deviation from TBM mixing (which usually predicts $\sin^2\theta_{12}=1/3$, 
$\sin^2\theta_{23}=1/2$ and $\theta_{13}=0$)  via Planck scale suppressed operators appearing solely  in the neutrino sector. 
The present construction (based on $A_4$ discrete symmetry) is such that observed $\theta_{13}$ can be successfully generated
in presence of such operators and deviations in $\theta_{12}$ and $\theta_{23}$ from their respective TBM values also arise at the same time. Such deviation 
from TBM mixing can be obtained via a unitary rotation (in the 13 plane) matrix, $U_{13}$, parameterised only by two parameters
$\theta$ and $\psi$ which also satisfy the condition $|(U_{13})_{11}|^2+|(U_{13})_{13}|^2=1$. The  correlations among $\theta$,
$\psi$ and the mixing angles are generic features of this class of models~\cite{King:2011zj, Altarelli:2012bn, Hernandez:2012ra}
considered here and can be realised very easily using $A_4$ discrete symmetry~\cite{Karmakar:2014dva, Karmakar:2015jza, 
Karmakar:2016cvb, Bhattacharya:2016rqj}. However in contrast
to all other models mentioned above, the present scenario predicts the Dirac CP phase $\delta$ to be maximal (as $\psi$ acquires
values 0 and $\pi$). Such predictions naturally make the model testable as well as distinguishable from other scenarios.
Now, the neutrino mass eigenvalues can be written as 
\begin{eqnarray}
 m_{1}&=&v^2x_3/|a|\sqrt{1+\alpha^2+2\alpha\cos\phi_{ba}}\label{m1}\\
m_{2}&=&v^2x_4/|a| \label{m2}\\
m_{3}&=&v^2x_3/|a|\sqrt{1+\alpha^2-2\alpha\cos\phi_{ba}}\label{m3}, 
\end{eqnarray}
where $\alpha=|b|/|a|$ and $\phi_{ba}=\phi_{b}-\phi_{a}$ is the phase difference between $b$ and $a$. 
Using these, one can now define a ratio of solar to atmospheric 
mass-squared differences ($\Delta{m}_{\odot}^{2}=\Delta{m_{21}^{2}}={m_{2}^{2}-m_{1}^{2}}$
and $|\Delta{m}_{A}^{2}|=|\Delta{m}^2_{31}|={m_{3}^{2}-m_{1}^{2}}
\approx|\Delta{m}^2_{32}|={m_{3}^{2}-m_{2}^{2}}$ 
respectively) defined by 
\begin{eqnarray}\label{r}
 r=\frac{\Delta{m}_{\odot}^{2}}{|\Delta{m}_{A}^{2}|}
 =\frac{(x_4^2(\alpha^2+2\alpha\cos\phi_{ba})-x_3^2)(1+\alpha^2-2\alpha\cos\phi_{ba})}{\pm x_3^24\alpha\cos\phi_{ba}}. 
\end{eqnarray}
Here, depending on the sign of denominator, the parameter space is divided into
two parts: '+' in the denominator yields normal hierarchy of neutrino mass whereas for '-' in the denominator yields inverted neutrino mass hierarchy. Therefore, the neutrino mixing angles and the ratio of solar to atmospheric mass-squared differences are altogether functions of five parameters, namely three coupling constants $y_1,y_2,y_{3}$ and $\alpha,\phi_{ba}$. Using the observed neutrino oscillation data summarised in latest global fit~\cite{Esteban:2018azc}, these parameters can be constrained  for both 
normal and inverted neutrino mass hierarchy, as we discuss below.

\subsection{Normal Hierarchy}
\begin{figure}[h]
$$
\includegraphics[height=5.5cm]{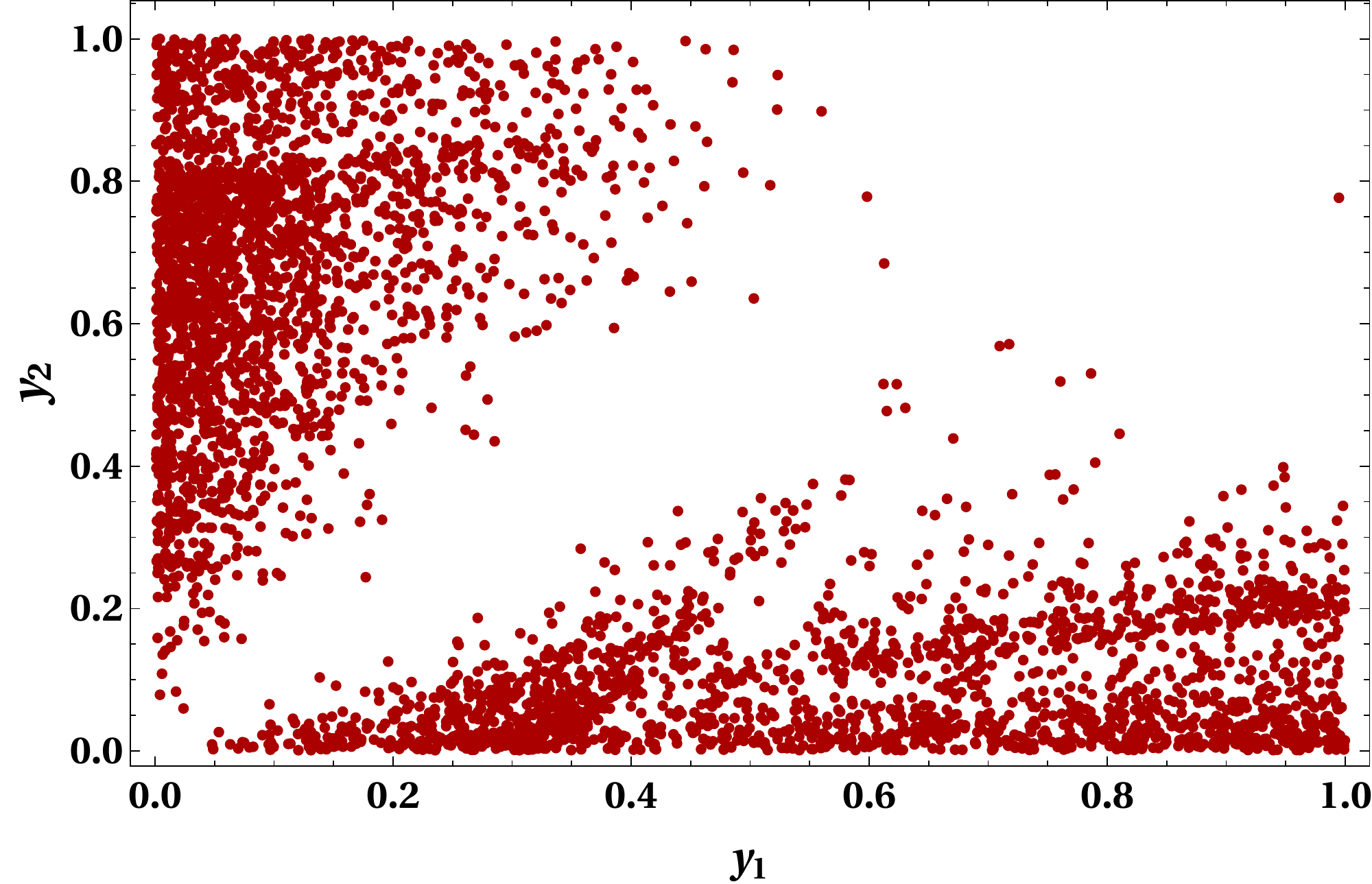}
\includegraphics[height=5.5cm]{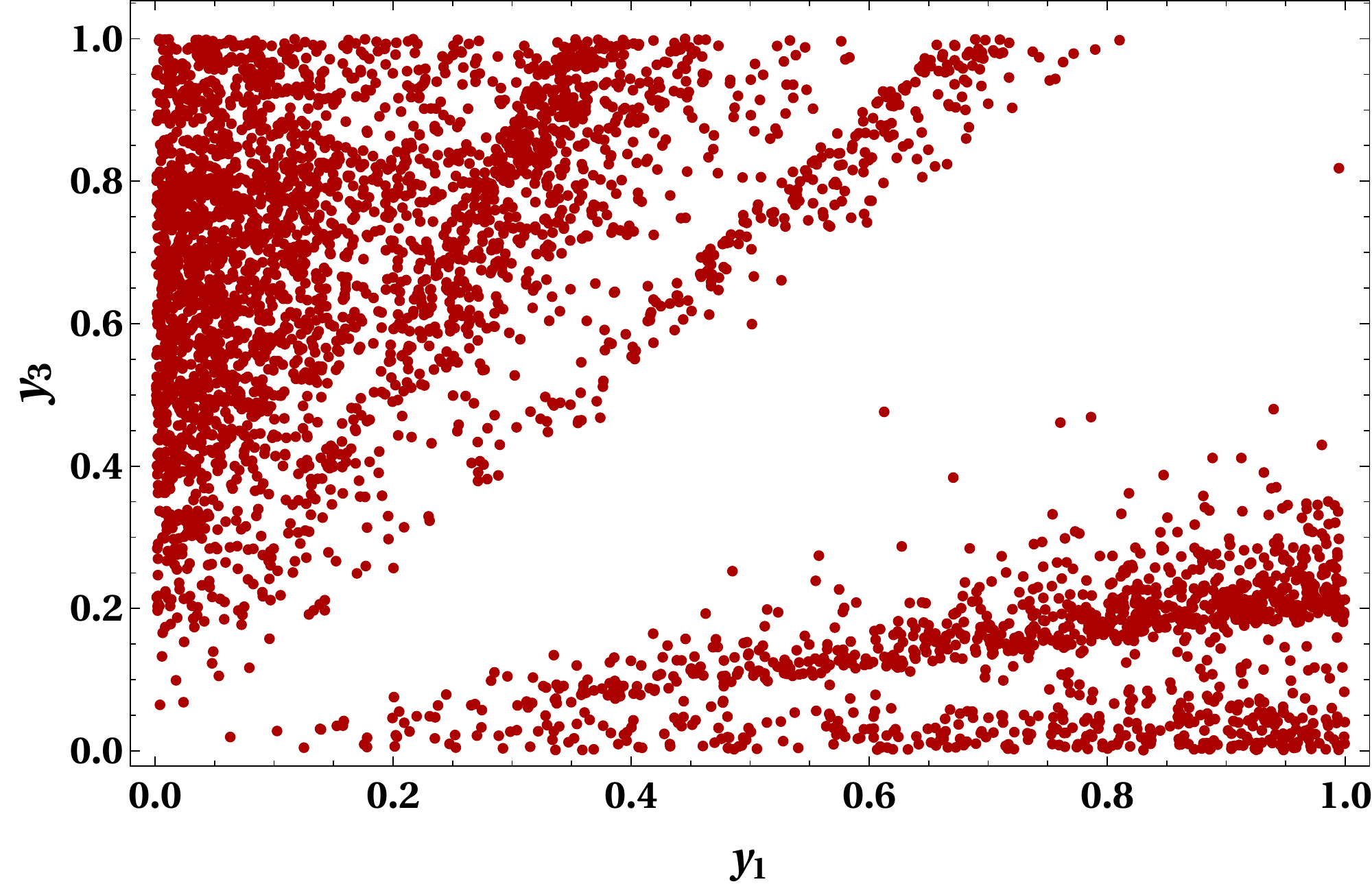}
$$
\caption{$y_{1}$ vs $y_{2}$ and $y_{1}$ vs $y_{3}$ satisfying correct neutrino data for normal hierarchy.}
\label{fig:nhy1y2}
\end{figure}
Here we first explore neutrino phenomenology for normal neutrino mass hierarchy. 
In left panel (right panel) of figure \ref{fig:nhy1y2} we have shown the
\begin{figure}[h]
$$
\includegraphics[height=5.5cm]{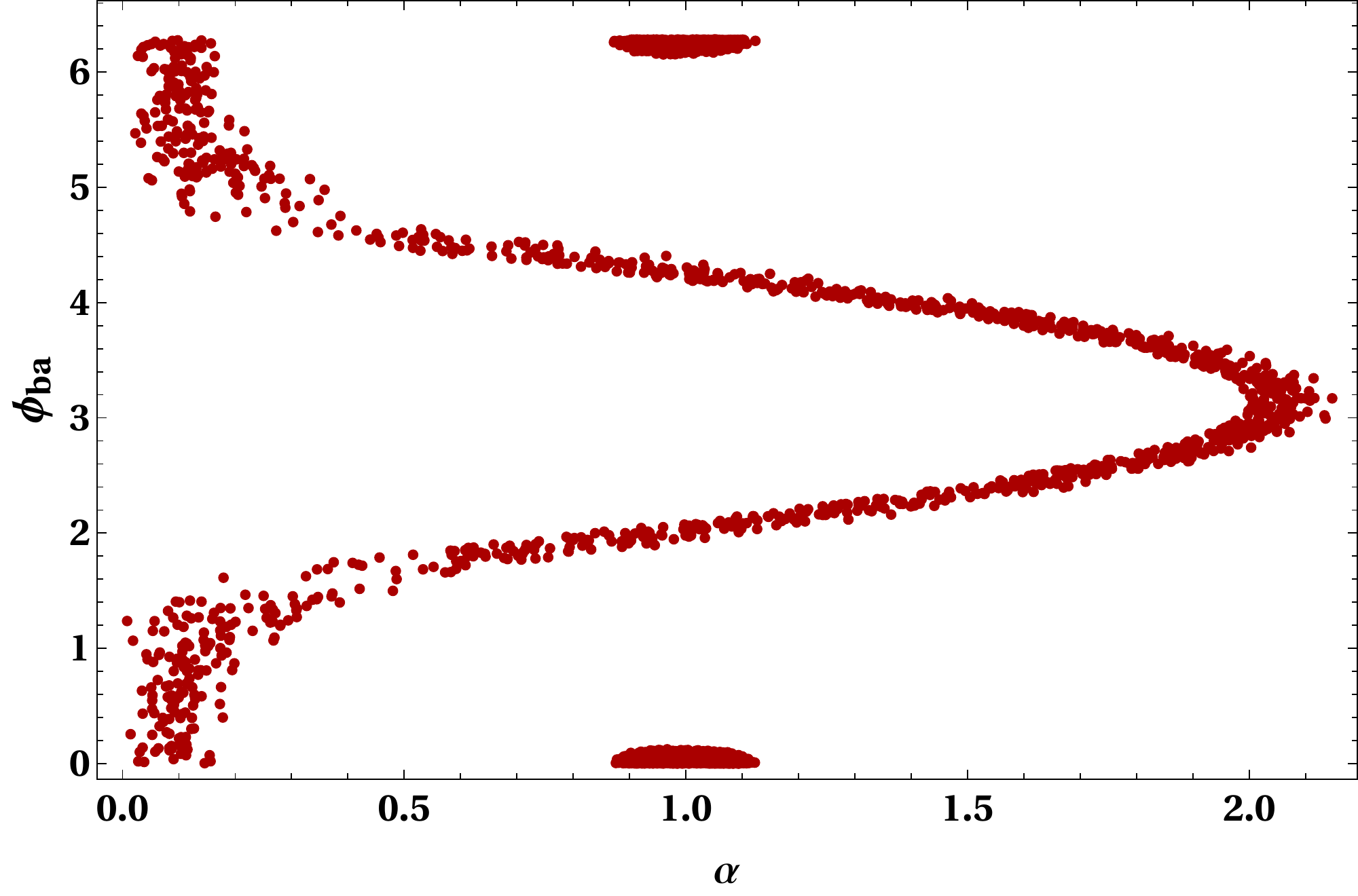}
\includegraphics[height=5.5cm]{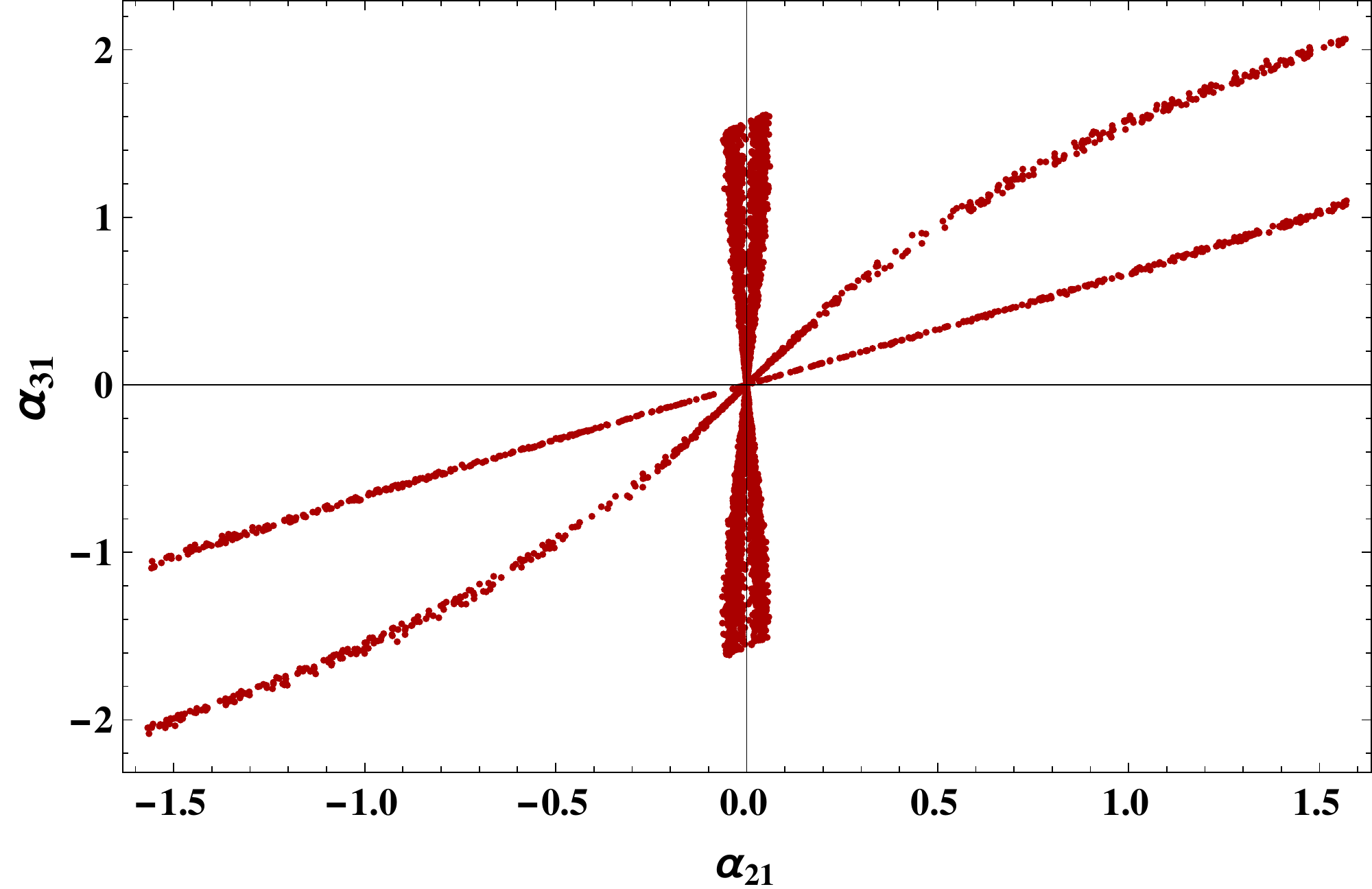}
$$
\caption{Left panel represents $\alpha$ vs $\phi_{ba}$ satisfying correct neutrino data. Right panel represents predicted regions for the Majorana phases $\alpha_{21}$ and $\alpha_{31}$ for normal hierarchy. }
\label{fig:nhapba}
\end{figure}
allowed points for $y_1$ vs $y_2$ ($y_3$). All these points satisfy 3$\sigma$ allowed range for the three neutrino mixing angles, 
ratio of solar to atmospheric mass-squared differences~\cite{Esteban:2018azc}. From the equation (\ref{m1}) and equation (\ref{m2}), one can easily obtain 
the common factor $|a|$ appearing in the absolute neutrino mass using $\Delta{m_{21}^{2}}={m_{2}^{2}-m_{1}^{2}}=2.43\times 10^{-5}$~eV$^2$.
Therefore the parameter space can be further constrained using the cosmological upper limit on the sum of absolute neutrino masses $\sum_i \lvert m_i \rvert < 0.11$ eV \cite{Aghanim:2018eyx}. 
\begin{figure}[h]
$$
\includegraphics[height=5.5cm]{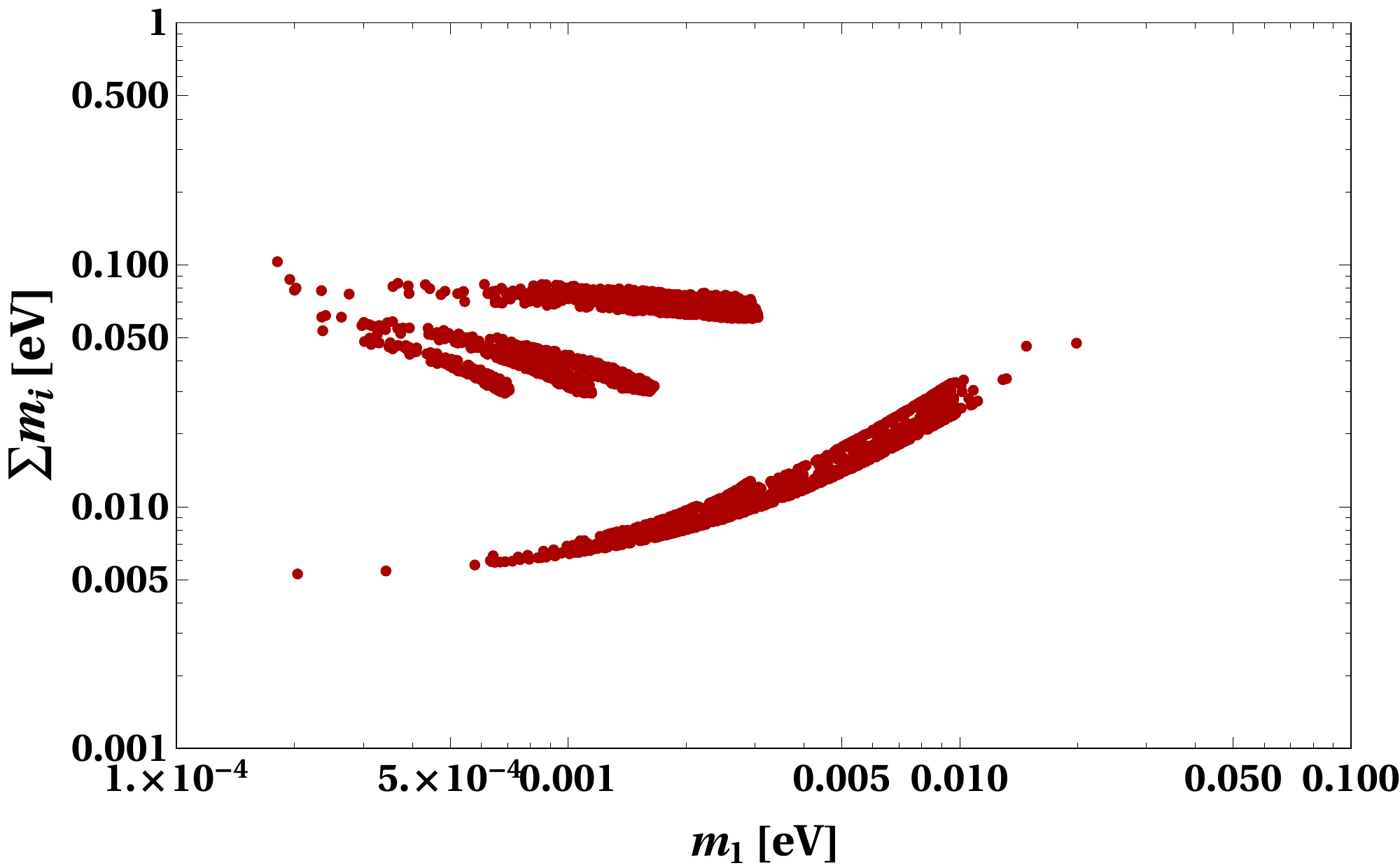}
\includegraphics[height=5.5cm]{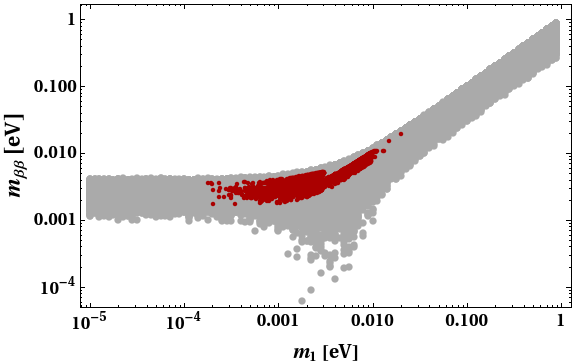}
$$
\caption{Predictions for lightest neutrino mass ($m_1$ for normal hierarchy) versus 
sum of absolute neutrino masses (left panel).  Predictions for lightest neutrino mass ($m_1$ for normal hierarchy) versus effective mass parameter for neutrinoless double beta decay (right panel).}
\label{fig:nhmass}
\end{figure}
The allowed regions presented in figure \ref{fig:nhy1y2} also satisfy this constraint. Now, in figure \ref{fig:nhapba} left 
panel we have plotted the allowed regions in the $\alpha$-$\phi_{ba}$ plane. As mentioned earlier here $\alpha$ is the ratio
of the two parameters appearing in the right handed neutrino mass matrix and $\phi_{ba}$ is their relative phase difference. This complex phase factor plays a crucial role in the present analysis in 
determination of the neutrino masses and the associated (Majorana) phases. 
\begin{figure}[h]
$$
\includegraphics[height=5.5cm]{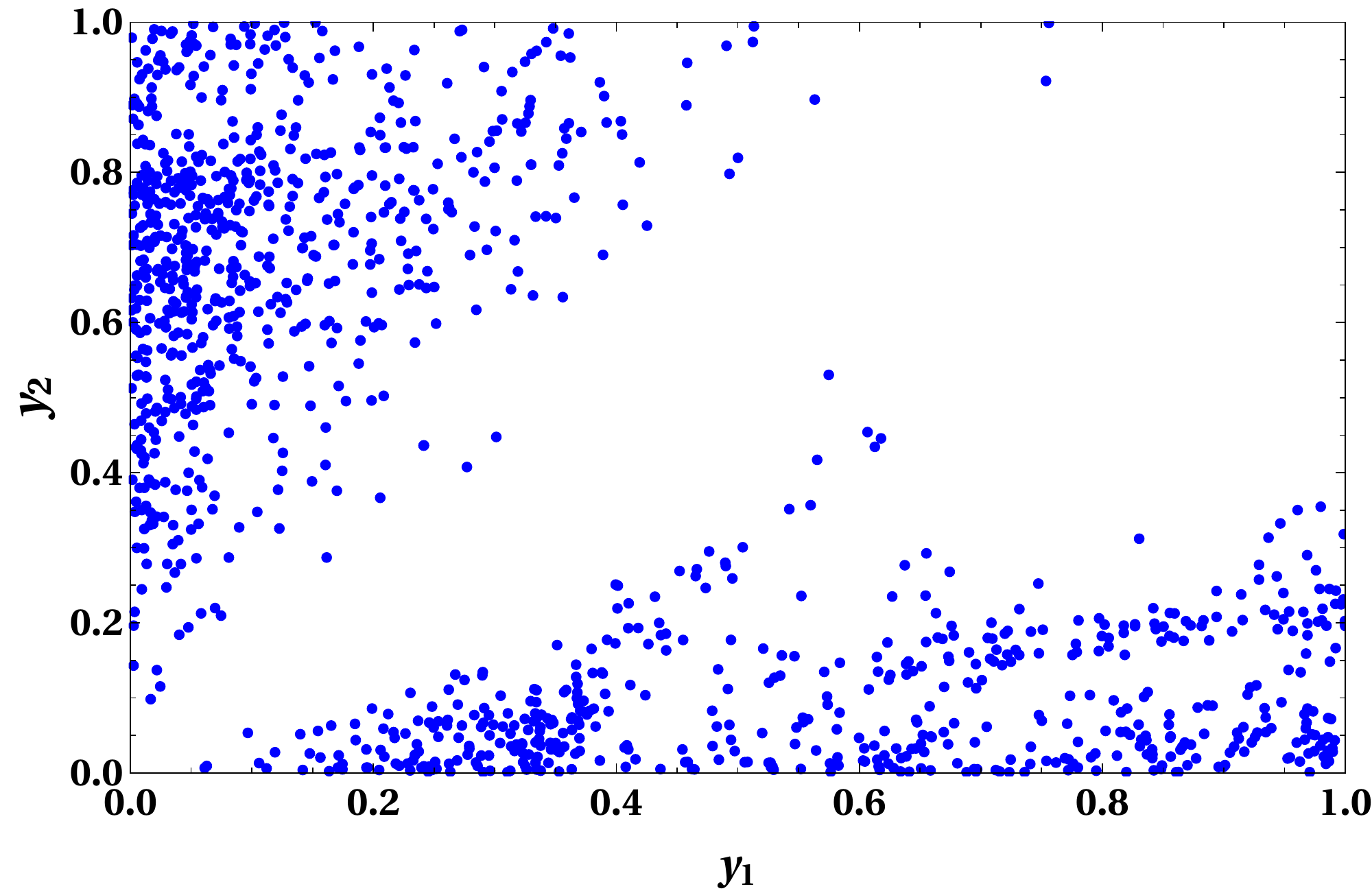}
\includegraphics[height=5.5cm]{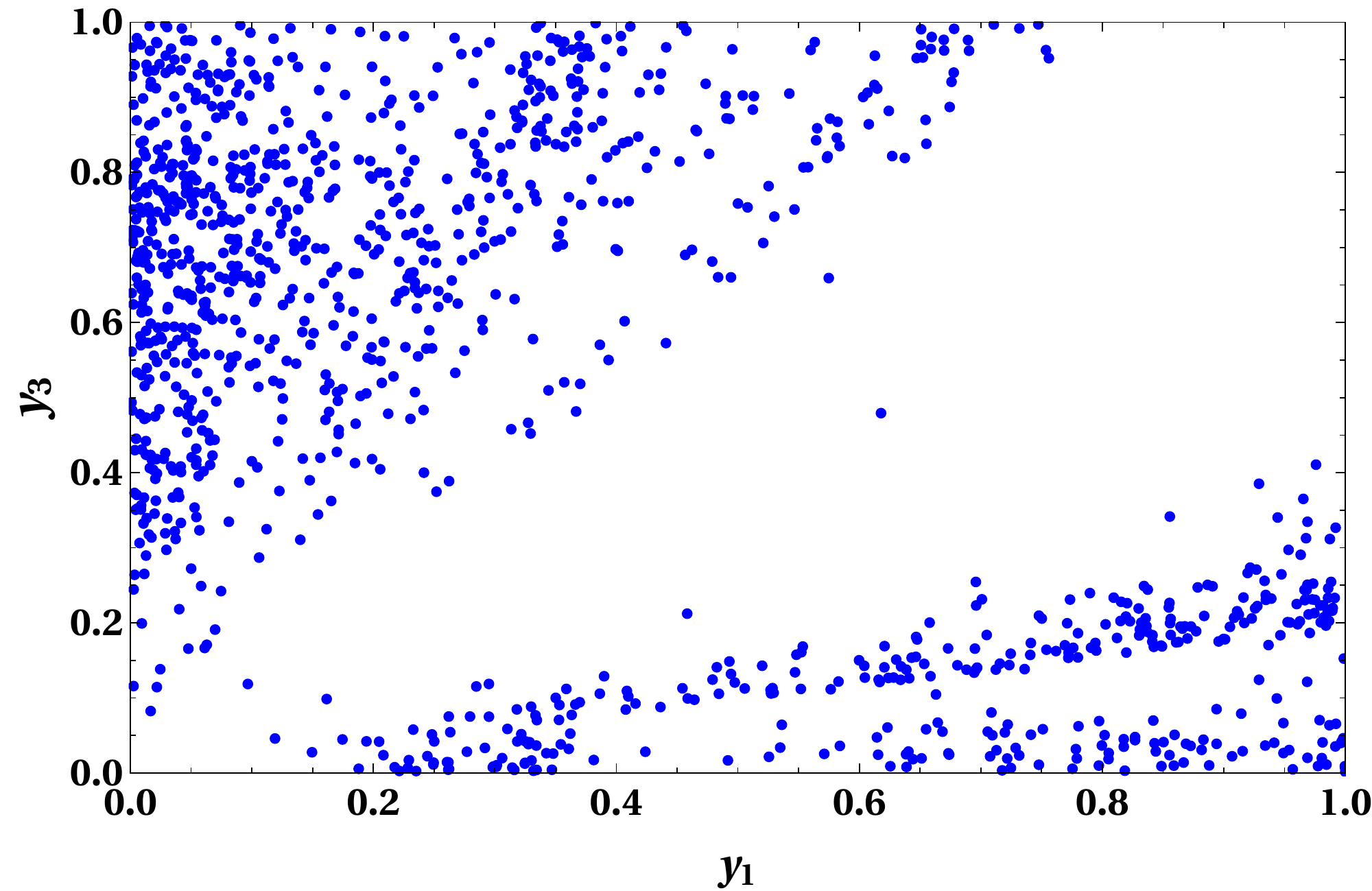}
$$
\caption{$y_{1}$ vs $y_{2}$ and $y_{1}$ vs $y_{3}$ satisfying correct neutrino data for inverted hierarchy.}
\label{fig:ihy1y2}
\end{figure}
%
In the right panel of figure \ref{fig:nhapba} we have shown the predictions for the Majorana phases in the current set-up.  
Predictions for the Majorana phases are also important to evaluate the absolute neutrino mass parameter ($m_{\beta\beta}$) appearing in the 
neutrinoless double beta decay amplitude, which is shown in the right panel of figure \ref{fig:nhmass}. In this figure the grey band represents
the standard light neutrino contribution to $m_{\beta\beta}$ for 3$\sigma$ range of neutrino oscillation parameters in case of normal hierarchy and the dark red points superimposed over the grey band are the predicted regions in our set-up. Here we find that the predicted region entirely falls inside the correct allowed region for  $m_{\beta\beta}$ in the standard three neutrino picture. 
In the left panel of figure \ref{fig:nhmass} we present the prediction regarding sum of absolute neutrino masses as a function 
of lightest light neutrino $m_1$ for normal hierarchy. As can be seen here, there are enough points lying within the cosmological upper bound on sum of absolute neutrino masses mentioned earlier.
%
\begin{figure}[h]
$$
\includegraphics[height=5.5cm]{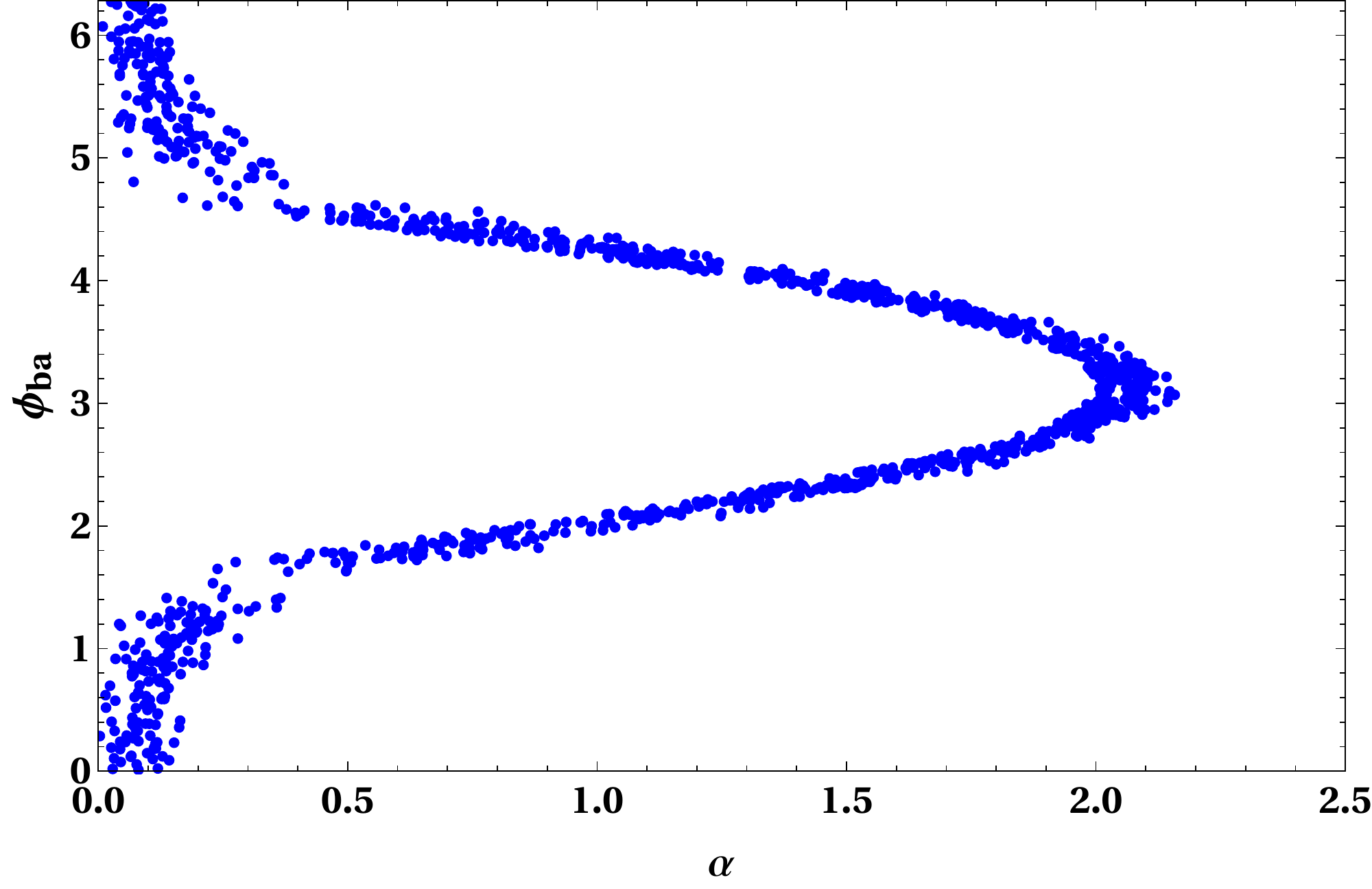}
\includegraphics[height=5.5cm]{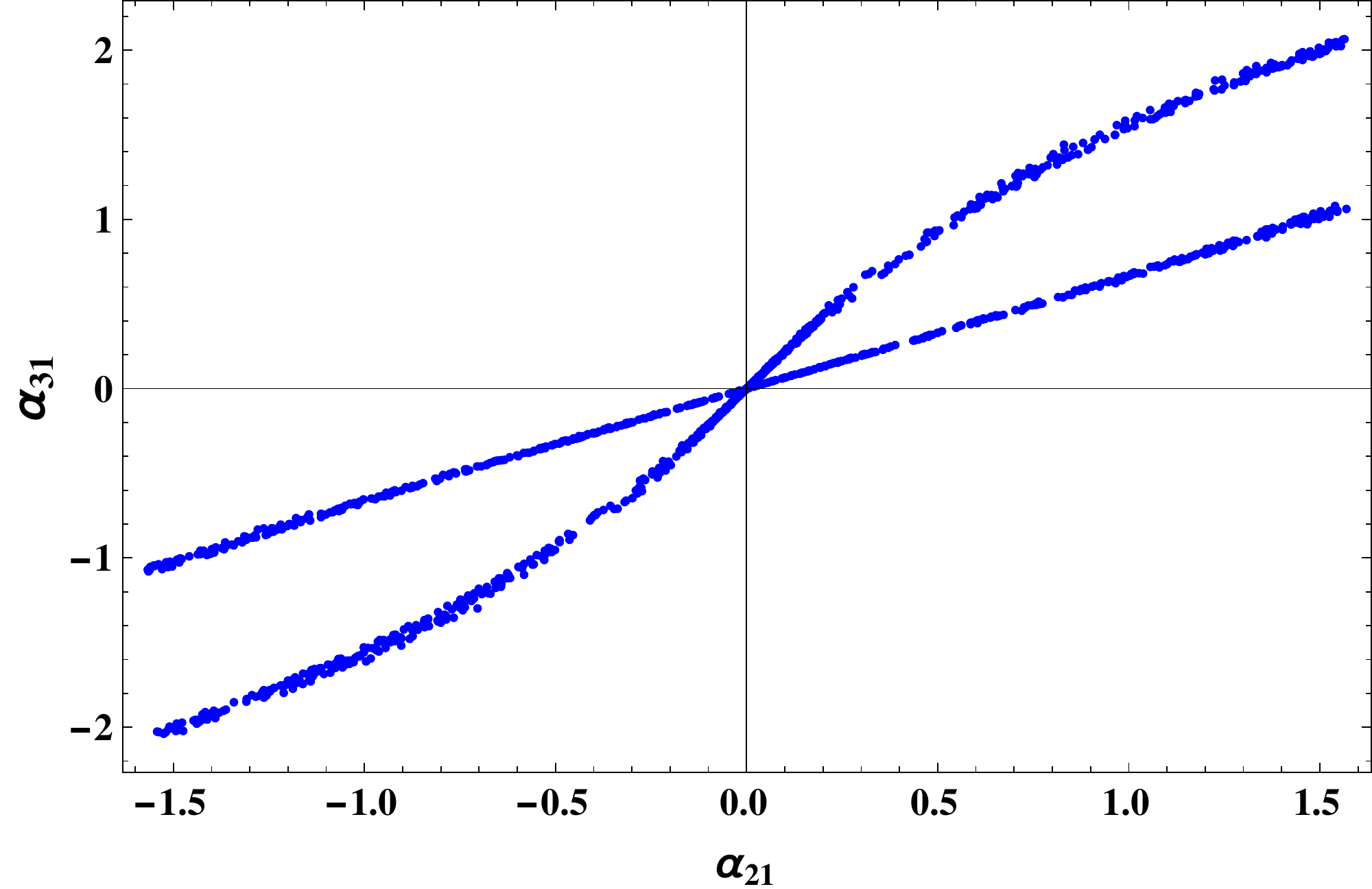}
$$
\caption{Left panel represents $\alpha$ vs $\phi_{ba}$ satisfying correct neutrino data. Right panel represents predicted regions for the Majorana phases $\alpha_{21}$ and $\alpha_{31}$ for inverted hierarchy.}
\label{fig:ihapba}
\end{figure}
%
\subsection{Inverted Hierarchy}
\begin{figure}[h]
$$
\includegraphics[height=5.5cm]{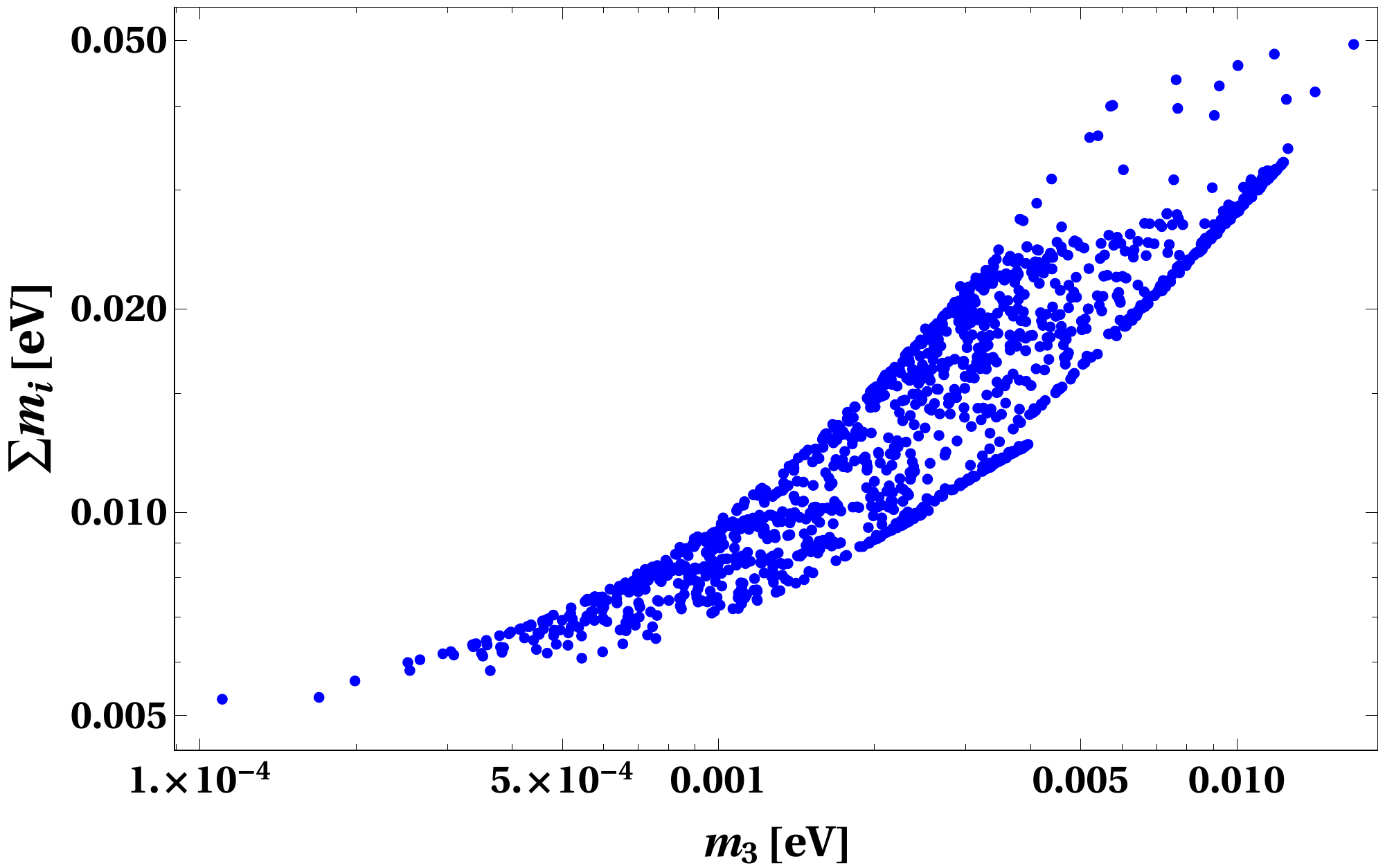}
\includegraphics[height=5.5cm]{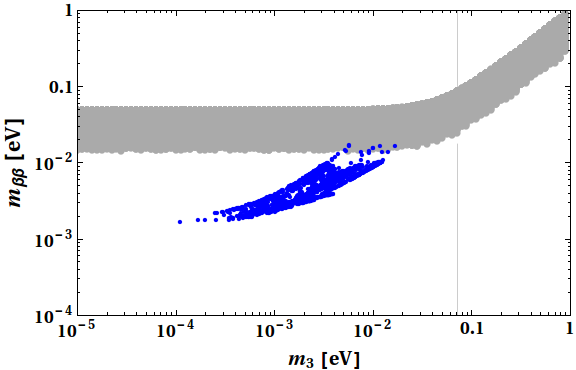}
$$
\caption{Predictions for lightest neutrino mass ($m_3$ for inverted hierarchy) versus 
sum of absolute neutrino masses (left panel). Predictions for lightest neutrino mass ($m_3$ for inverted hierarchy) versus effective mass parameter for neutrinoless double beta decay (right panel).}
\label{fig:ihmass}
\end{figure}

Using a similar strategy, we analyse the allowed parameter space for inverted neutrino mass hierarchy. In figure \ref{fig:ihy1y2}, the blue dots in 
left panel (right panel) represent allowed points for $y_1$ vs $y_2$ ($y_3$) satisfying correct 
neutrino oscillation data. In figure \ref{fig:ihapba}, left panel shows allowed points in the $\alpha$ - $\phi_{ba}$ plane
whereas the right panel shows the predictions for the two Majorana phases $\alpha_{21}$  and $\alpha_{31}$ for 
inverted neutrino mass hierarchy. 
Finally in figure \ref{fig:ihmass} we show the predictions for absolute neutrino masses (left panel) and $m_{\beta\beta}$ as a
function of lightest neutrino mass $m_3$. Interestingly, for values $y_{1,2,3}\leq 1$, we find that the predictions for the 
effective neutrino mass parameter $m_{\beta\beta}$ mostly fall outside of the standard allowed regions (shown as grey coloured band) for inverted hierarchy. This makes the 
inverted hierarchy in the current scenario less favourable compared to the normal hierarchy. However, such conclusions are based on the very specific Planck suppressed corrections we have considered in our analysis and they will change if more corrections are included. However, the essence of the set-up is that the correct light neutrino phenomenology can be obtained for different subsets of the entire parameter space.

It should be noted that the Planck scale suppressed corrections crucially depend upon the ratio $u/M_{\rm Pl}$ up to some Yukawa couplings. In the absence of any fine tuning of Yukawa couplings, all the corrections to neutrino sector go as $M_R/M_{\rm Pl}$ since $M_R \propto u$. In fact the usual Weinberg operator $(L L H H)/M_{\rm Pl}$ can give a correction like $\frac{M_R}{M_{\rm Pl}} M_{\nu}$ where $M_{\nu}$ is the leading order type I seesaw contribution to light neutrino mass. For example, if the scale of $A_4$ breaking at renormalisable level is $u \sim 10^{16}$ GeV, similar to typical grand unified theory (GUT) scale, one can have a slightly lower scale $M_R \sim 10^{14}$ GeV and hence the scale $a, b$ in the right handed neutrino mass matrix given in \eqref{mright}, by appropriate tuning of relevant Yukawa couplings. Such $M_R$ can generate sub-eV scale light neutrino mass matrix at leading order for order one Dirac Yukawa couplings. If we consider the corrections to Dirac neutrino mass matrix, such a scale would correspond to $y_2, y_3 \leq 10^{-3}-10^{-2}$ which will correspond to the lowermost regions (along the y-axis) of the parameter space shown in figure \ref{fig:nhy1y2} and figure \ref{fig:ihy1y2} described before. Similar estimates about these couplings can be made for different symmetry breaking scale. However, in order to have sizeable corrections to the TBM mixing and to avoid unnaturally fine-tuned as well as non-perturbative Yukawa couplings, the $A_4$ scale should be close to GUT scale. As far as fine-tuning is concerned, we set a tolerance same as electron Yukawa in the standard model. Different $A_4$ breaking scale will also have different implications for the super-WIMP dark matter sector, requiring some amount of fine tuning in relevant Yukawa couplings related to mother particle and dark matter, as we discuss below.

\section{Super-WIMP Dark Matter}
\label{sec3}
The dark matter in our setup is similar to the super-WIMP scenario \cite{Feng:2003uy} where a metastable WIMP decays into super-weakly interacting dark matter at late epochs. As can be seen from the dimension five Lagrangian \eqref{plH}, there are several terms which can give rise to decay of either $\psi$ or $\eta$, the particles of the $Z_4$ sector. At renormalisable level however, both of them are stable due to the unbroken $Z_4$ symmetry. While $\eta$ can be produced thermally in the early universe due to its electroweak gauge interactions, the singlet fermion $\psi$ has negligible thermal abundance. In a general super-WIMP formalism where $\eta$ is the metastable WIMP and $\psi$ is the DM candidate, one can write down the the decay width of $\eta$ into two dark matter particles $(\psi)$ as 
\begin{equation}
\Gamma_{\eta \rightarrow \psi \bar{\psi}}=\frac{Y^2_{\eta \psi}(m_{\eta}^2-4m_{\psi}^2)}{8\pi m_\eta}\sqrt{1-\frac{4m_{\psi}^2}{m_{\eta}^2}}
\label{decay:psipsi}
\end{equation}
where $Y_{\eta \psi}$ is the effective Yukawa coupling, m$_\eta$ and m$_\psi$ are the masses of the mother particle $\eta$ and $\psi$ respectively. 

However, this doublet scalar, apart from having electroweak gauge interactions, can have sizeable quartic interactions with other scalars like the standard model Higgs doublet and hence can be thermally produced in the early universe. Now, considering the mother particle $\eta$ to be in thermal equilibrium in the early universe which also decays into the dark matter particle $\psi$, we can write down the relevant Boltzmann equations for comoving number densities of $\eta, \psi$ as
\begin{eqnarray}
\frac{dY_{\eta}}{dx} &=& -\frac{4 \pi^2}{45}\frac{M_{\rm Pl} M_{\rm sc}}{1.66}\frac{\sqrt{g_{\star}(x)}}{x^2}\Bigg[\sum_{p\,=\,\text{SM\, particles}}
\langle\sigma {\rm v} \rangle_{\eta \eta \rightarrow p\bar{p}}\left( Y_{\eta}^2-(Y_{\eta}^{\rm eq})^2\right) \Bigg]\nonumber \\
&& - \frac{M_{\rm Pl}}{1.66} \frac{x\sqrt{g_{\star}(x)}}{M_{\rm sc}^2\ g_s(x)} \Gamma_{\eta \rightarrow \bar{\psi} \psi}  \ Y_{\eta},
\label{BEeta}
\end{eqnarray}
\begin{eqnarray}
\frac{dY_\psi}{dx} &=& \frac{2 M_{\rm Pl}}{1.66} \frac{x\sqrt{g_{\star}(x)}}{M_{\rm sc}^2 \ g_s(x)} \Gamma_{\eta \rightarrow \bar{\psi} \psi}
\ Y_{\eta}
\label{BEpsi}
\end{eqnarray}
where $x=\dfrac{M_{\rm sc}}{T}$, is a dimensionless variable while
$M_{\rm sc}$ is some arbitrary mass scale which we choose equal to
the mass of $\eta$ and $M_{\rm Pl}$ is the Planck mass. Moreover, $g_s(x)$
is the number of effective degrees of freedom associated to the
entropy density of the universe and the quantity $g_{\star}(x)$
is defined as
\begin{eqnarray}
\sqrt{g_{\star}(x)} = \dfrac{g_{\rm s}(x)}
{\sqrt{g_{\rho}(x)}}\,\left(1 -\dfrac{1}{3}
\dfrac{{\rm d}\,{\rm ln}\,g_{\rm s}(x)}{{\rm d} \,{\rm ln} x}\right)\,. 
\end{eqnarray}
Here, $g_{\rho}(x)$ denotes the effective number of degrees
of freedom related to the energy density of the universe at
$x=\dfrac{M_{\rm sc}}{T}$. 
The first term on the right hand side of the Boltzmann equation \eqref{BEeta} corresponds to the self annihilation of $\eta$ into standard model particles and vice versa which play the role in its freeze-out. The second term on the right hand side of this equation corresponds to the dilution of $\eta$ due to its decay into dark matter $\psi$. Let us denote the freeze-out temperature of $\eta$ as $T_F$ and its decay temperature as $T_D$. If we assume that the mother particle freezes out first followed by its decay into dark matter particles, we can consider $T_F > T_D$. In such a case, we can first solve the Boltzmann equation for $\eta$ considering only the self-annihilation part to calculate its freeze-out abundance. \begin{eqnarray}
\frac{dY_{\eta}}{dx} &=& -\frac{4 \pi^2}{45}\frac{M_{\rm Pl} M_{\rm sc}}{1.66}\frac{\sqrt{g_{\star}(x)}}{x^2}\Bigg[\sum_{p\,=\,\text{SM\, particles}}
\langle\sigma {\rm v} \rangle_{\eta \eta \rightarrow p\bar{p}}\left( Y_{\eta}^2-(Y_{\eta}^{\rm eq})^2\right) \Bigg]
\label{BEeta1}
\end{eqnarray}
Then we solve the following two equations for temperature $T < T_F$
\begin{eqnarray}
\frac{dY_{\eta}}{dx} &=&  - \frac{M_{\rm Pl}}{1.66} \frac{x\sqrt{g_{\star}(x)}}{M_{\rm sc}^2\ g_s(x)} \Gamma_{\eta \rightarrow \bar{\psi} \psi}  \ Y_{\eta},
\label{BEeta2}
\end{eqnarray}
\begin{eqnarray}
\frac{dY_\psi}{dx} &=& \frac{2 M_{\rm Pl}}{1.66} \frac{x\sqrt{g_{\star}(x)}}{M_{\rm sc}^2 \ g_s(x)} \Gamma_{\eta \rightarrow \bar{\psi} \psi}
\ Y_{\eta}.
\label{BEpsi2}
\end{eqnarray}
We stick to this simplified assumption $T_F > T_D$ in this work and postpone a more general analysis without any assumption to an upcoming work. The assumption $T_F > T_D$ allows us to solve the Boltzmann equation \eqref{BEeta1} for $\eta$ first, calculate its freeze-out abundance and then solve the corresponding equations \eqref{BEeta2}, \eqref{BEpsi2} for $\eta, \psi$ using the freeze-out abundance of $\eta$ as initial condition\footnote{Recently another scenario was proposed where the dark matter freezes out first with underproduced freeze-out abundance followed by the decay of a long lived particle into dark matter, filling the deficit \cite{Borah:2017dfn, Biswas:2018ybc}.}. In such a scenario, we can solve the Boltzmann equations \eqref{BEeta2}, \eqref{BEpsi2} for different benchmark choices of $Y, m_{\eta}, m_{\rm DM} \equiv M_{\psi}$ and estimate the freeze-out abundance of $\eta$ that can generate $\Omega h^2 =0.12$, the {\it canonical} value of the dark matter $(\psi)$ relic abundance in the present universe. This required freeze-out abundance of $\eta$ then restricts the parameters involved in its coupling to the SM particles. It turns out that a scalar singlet like $\eta$ interacts with the SM particles only through the Higgs portal and hence depends upon the $\eta-H$ coupling, denoted by $\lambda_{H \eta}$. To find out the freeze-out abundance of $\eta$, we have used \texttt{micrOMEGAs} package \cite{Belanger:2013oya} in our work.

Now, as can be seen from the Planck suppressed terms of \eqref{plH}, there are three different decay modes of $\eta$ namely, $\eta \rightarrow l \bar{l}, l \psi, \psi \psi$, where $l$ denotes a SM lepton. Out of these, the decay mode $\eta \rightarrow \psi \psi $ has effective coupling $\tilde{Y_1} \frac{v}{M_{\rm Pl}}$ where $v$ is the Higgs VEV at electroweak scale. The other two decay modes have larger effective coupling $\tilde{Y'_2} \frac{v_{\phi}}{M_{\rm Pl}}, \tilde{Y''_2} \frac{v_{\phi}}{M_{\rm Pl}}$ where $v_{\phi}$ is the $A_4$ flavon VEV. Since flavon VEV is much larger than electroweak one, the first two decay modes will be more dominant than the last one. Out of these two, the mode $\eta \rightarrow l \psi$ will contribute to the abundance of DM $\psi$. One can also think of some non-minimal scenarios where there exists some local symmetries which forbids $\eta \rightarrow l \bar{l}$ or even $\eta \rightarrow l \psi$ depending upon how $\eta, \psi$ are charged under the local symmetry. In the former case, DM will be produced dominantly from $\eta \rightarrow l \psi$ mode while in the second case the only possible mode that remains is the $\eta \rightarrow \psi \psi$ one. Since these two modes have very different effective couplings, the DM phenomenology can be very different. We now discuss all these three distinguishing cases one by one.

We note that our second and third cases will involve some local symmetries which are not explicitly broken by Planck scale physics. Here we do not outline a complete model for such cases but point out the distinguishing DM phenomenology without affecting the neutrino phenomenology. The additional gauge interactions of $\psi, \eta$ which can prevent coupling of $\psi, \eta$ to SM leptons can also produce $\psi$ thermally in the early universe, in principle. However, if such mediator gauge bosons are very heavy, with masses greater than the reheat temperature after inflation, then the thermal production of such DM remains suppressed \cite{Mambrini:2013iaa} and non thermal contribution from the metastable WIMP will become more relevant. Another way is to assume very feeble gauge interactions so that they never get thermalised.

\subsection{Case I}
In this subsection we discuss the DM phenomenology in the most general way, corresponding to the symmetries and particle content of the $A_4 \times Z_4$ model discussed before. As mentioned earlier, the Planck scale suppressed operators open up several decay modes of the $Z_4$ sector particles namely, $\eta, \psi$. The scalar doublet $\eta$, the metastable WIMP in our case can have three different decay modes out of which the decay into a pair of leptons and a lepton plus DM are the dominant ones. The corresponding decay widths can be written as 
\begin{equation}
\Gamma_{\eta \rightarrow e^{-}e^{+}}=\frac{\lambda_{1}^2(m_{\eta}^2-4m_{e}^2)}{8\pi m_\eta}\sqrt{1-\frac{4m_{e}^2}{m_{\eta}^2}}
\label{decay:epem}
\end{equation}

\begin{equation}
\Gamma_{\eta \rightarrow \nu \psi}=\frac{\lambda_{2}^2\bigg(m_{\eta}^2-m_{\psi}^2-m_{\nu}^2-2m_{\psi} m_{\nu}\bigg)}{8\pi m_\eta}\sqrt{(1-\frac{(m_{\psi}-m_{\nu})^2}{m_{\eta}^2})(1-\frac{(m_{\psi}+m_{\nu})^2}{m_{\eta}^2})}
\label{decay:nupsi}
\end{equation}
where, $\lambda_{1}$ and $\lambda_{2}$ can be written as $\frac{Y_{1}^{\prime} \langle \phi_{N_{i}} \rangle}{M_{Pl}}$ and $\frac{Y_{2}^{\prime \prime}  \langle \phi_{N_{i}} \rangle}{M_{Pl}}$ respectively.

However in this case, DM is also not perfectly stable as the $Z_4$ symmetry which protects its stability gets explicitly broken by Planck suppressed terms. As can be seen from \eqref{plH}, $\psi$ can decay into the Higgs boson and leptons due to the term $Y^{''}_1 \overline{L_{\alpha}} \tilde{H} \psi$. To forbid this decay at tree level, we consider the DM mass to be below 1 MeV. This term also can give rise to a mixing of $\psi$ with light neutrinos of the order 
\begin{equation}
\sin{2\theta} \approx Y{''}_1 \frac{ \langle H \rangle \langle \phi_N \rangle}{M_{\rm Pl} m_{\psi}}
\label{mixing1}
\end{equation}
which can be tuned appropriately in order to satisfy X-ray or gamma ray bounds.

\begin{figure}[h!]
\centering
\includegraphics[width=0.6\textwidth]{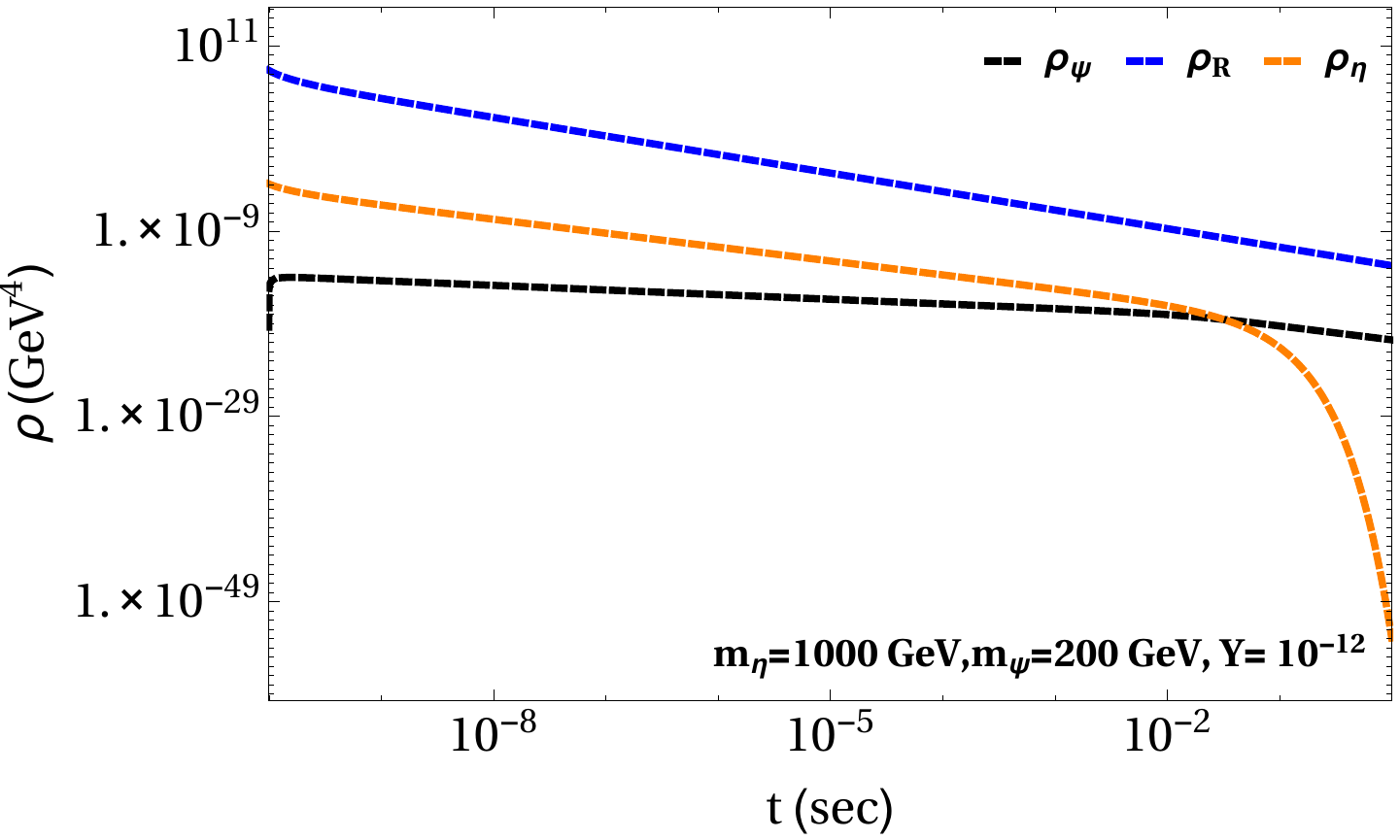}
\caption{The evolution of $\eta, \psi \equiv {\rm DM}$ number density and radiation energy density with time for case I.}
\label{entropy}
\end{figure}

In this case, the metastable WIMP $\eta$ not only decays to DM but also the charged lepton pairs. Though DM remains out of thermal equilibrium throughout, the charged leptons are part of the thermal bath and hence producing them from $\eta$ decay can, in principle, release entropy. In that case, solving the coupled Boltzmann equations for $\eta$ and $\psi$ described earlier is insufficient and one has to consider a third equation for radiation energy density of the universe. Before going to the actual DM calculations, we first check the amount of entropy release by solving the following three coupled Boltzmann equations namely,
\begin{eqnarray}\nonumber
\frac{dn_{\eta^0}}{dt} & =& -3H n_{\eta^0}-\Gamma_{\eta}n_{\eta^0}\\ 
\nonumber
\frac{d\rho_R}{dt} & =& -4H \rho_{R}+\Gamma_{\eta\rightarrow l^+ l^-}\rho_{\eta}\\
\frac{dn_\psi}{dt} & =& -3H n_{\psi}+\Gamma_{\eta\rightarrow \psi \nu}n_{\eta}
\label{radiation}
\end{eqnarray}
where $\Gamma_{\eta}=\Gamma_{\eta\rightarrow l^+ l^-} + \Gamma_{\eta\rightarrow \psi \nu}$. Please note that here we are writing the Boltzmann equations in terms of ordinary number density $(n)$, energy density $(\rho)$ and time $(t)$ unlike the earlier equations in terms of comoving quantities. 
The first equation above shows the time evolution of the number density of $\eta$ where the usual dilution due to the expansion of the universe is given by the first term on the right hand side (RHS) and dilution due to the decay is given by the second term on the RHS. There should, in principle, be one more term on RHS namely, $- \langle \sigma v \rangle_{\eta \eta \rightarrow pp} \bigg[ n_{\eta^0}^2-(n_{\eta^0}^{eq})^2\bigg]$ with $p$ denoting any SM particles to which $\eta$ can self annihilate into. However, we have dropped this term assuming that $\eta$ has already frozen out before it starts decaying which is a reasonable assumption (like super-WIMP framework) for the couplings governing $\eta$ decay in our model. The second and third equations are for the evolution of radiation energy density and the DM number density respectively. The figure \ref{entropy} shows that the contribution to the radiation or entropy release due to the decay of $\eta$ to the lepton pairs is negligible all the way up to the epochs where $\eta$ decays to reduce its number density while yielding $\psi$. The radiation energy density remains dominant during this and does not get any significant contribution from $\eta$ decay. Since the entropy release is negligible, it thereby justifies the use of coupled Boltzmann equations only for metastable WIMP and DM in our analysis.

\begin{figure}[h!]
		\includegraphics[width=0.45\textwidth]{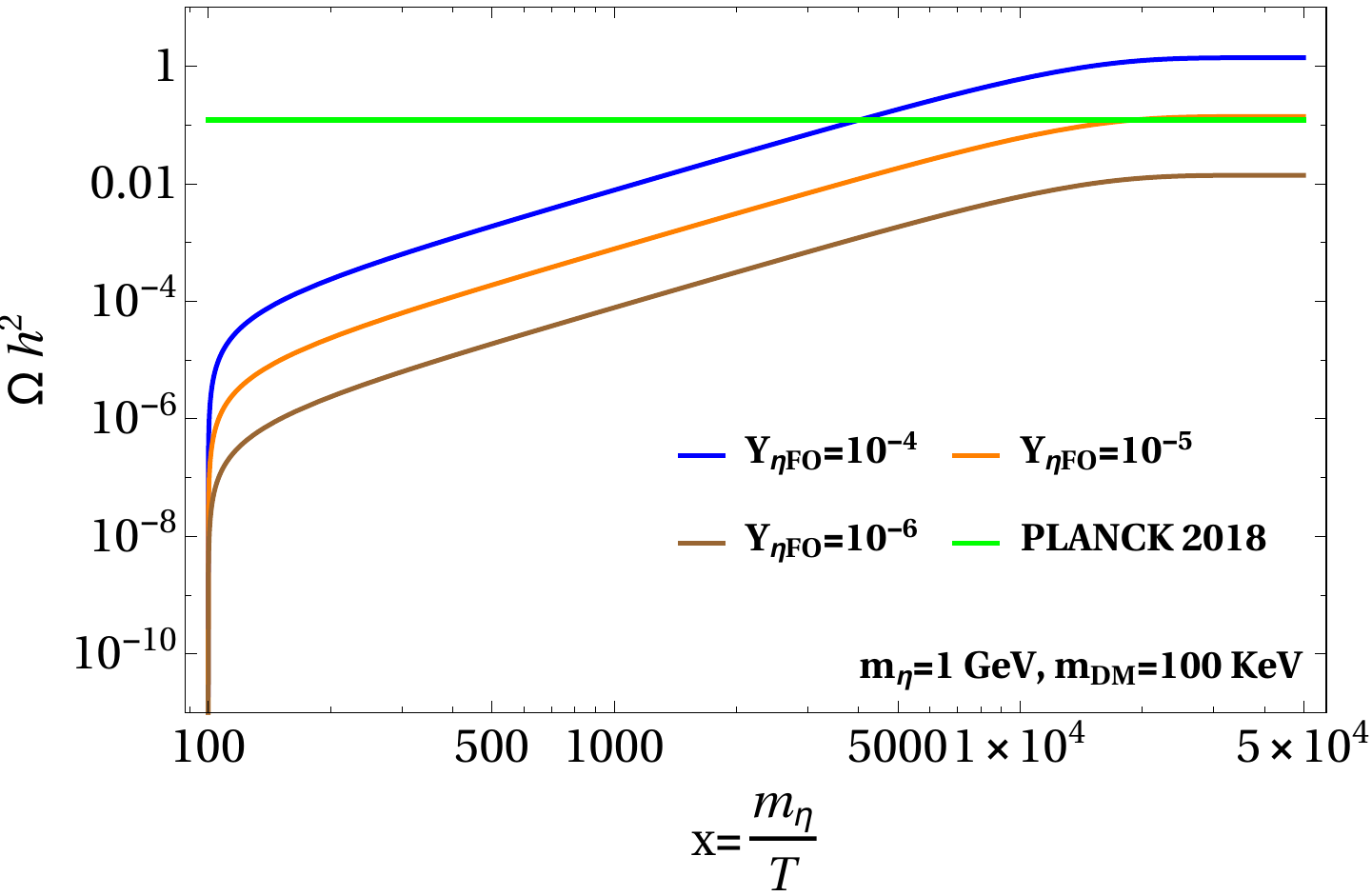}
		\includegraphics[width=0.45\textwidth]{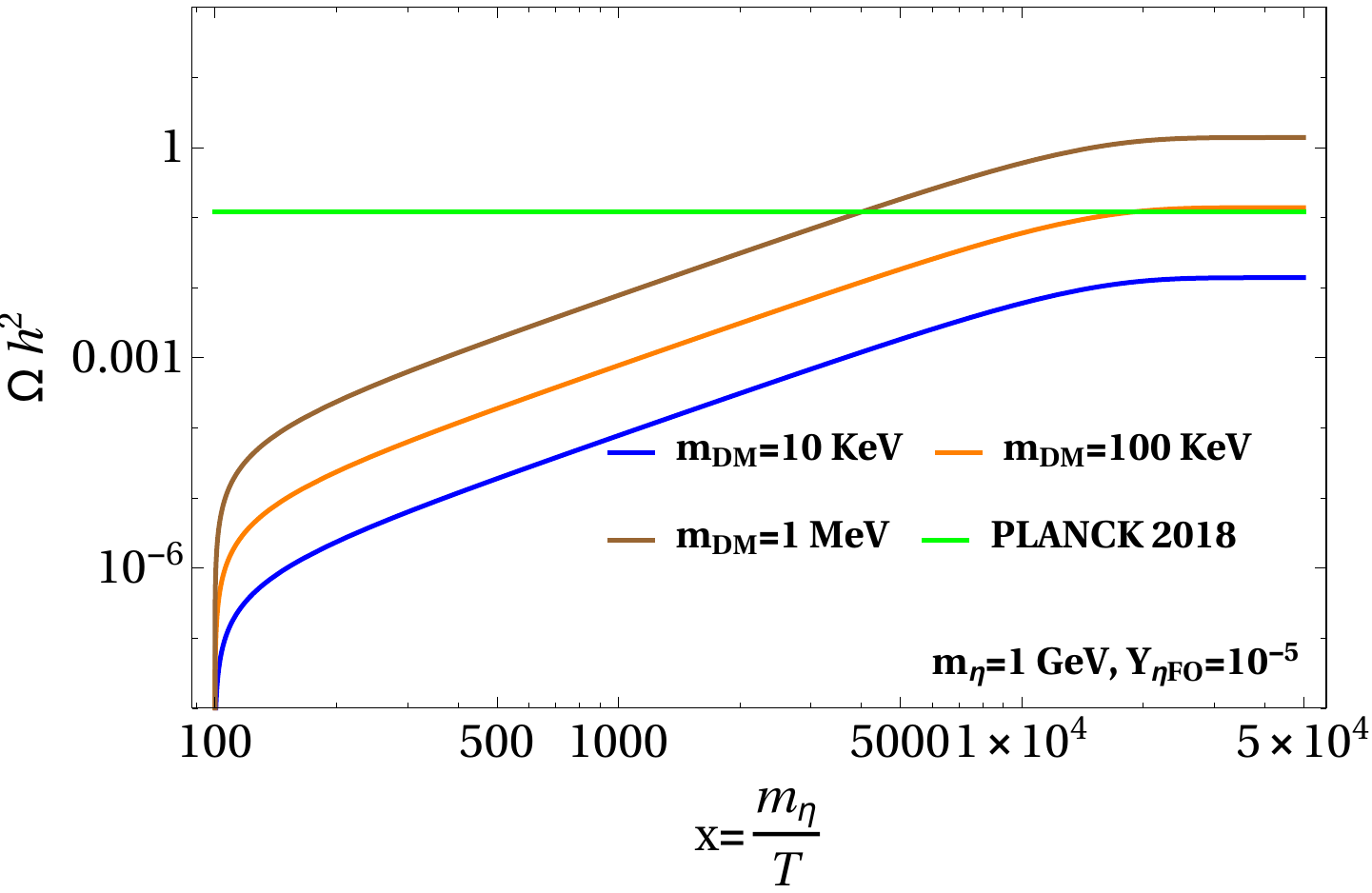} \\
		\includegraphics[width=0.45\textwidth]{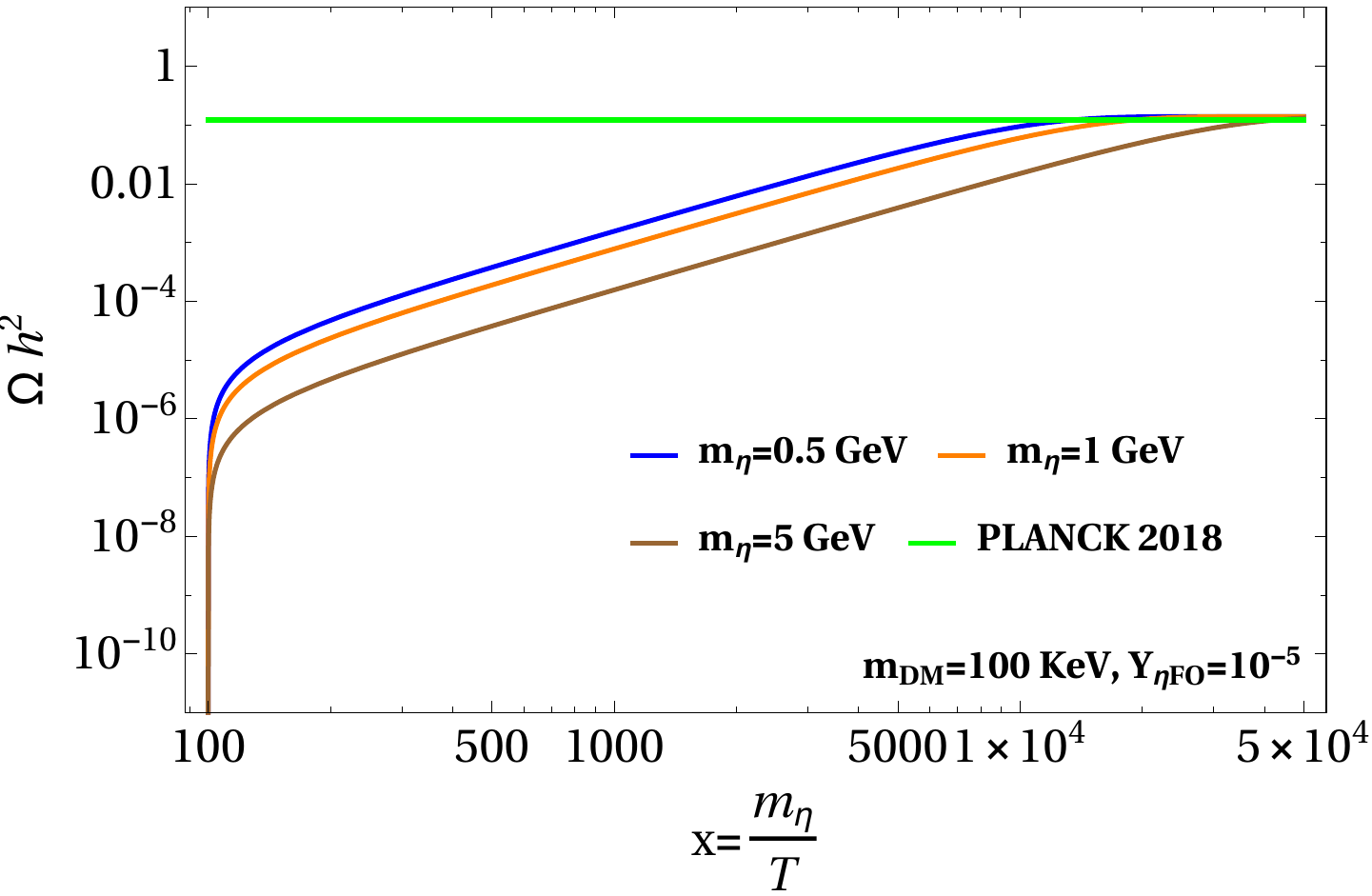}
		\caption{Comparison of $\Omega_{DM} h^2$ with respect to different model parameters  for case I.}
\label{Fig:compare-Y1}
\end{figure}

We now write down the coupled Boltzmann equations for $\eta, \psi$ in terms of their comoving number densities as
\begin{eqnarray}
\frac{dY_{\eta}}{dx} &=&  - \frac{M_{\rm Pl}}{1.66} \frac{x\sqrt{g_{\star}(x)}}{M_{\rm sc}^2\ g_s(x)} \Gamma_{\eta}  \ Y_{\eta},
\label{BEeta:case1}
\end{eqnarray}
\begin{eqnarray}
\frac{dY_\psi}{dx} &=& \frac{M_{\rm Pl}}{1.66} \frac{x\sqrt{g_{\star}(x)}}{M_{\rm sc}^2 \ g_s(x)} \Gamma_{\eta \rightarrow \psi \bar{\nu}}
\ Y_{\eta}.
\label{BEpsi:case1}
\end{eqnarray}
We solve these two equations \eqref{BEeta:case1} and \eqref{BEpsi:case1} for different benchmark values of the parameters and show the results in figure \ref{Fig:compare-Y1}. Since we are solving these equations after the freeze-out of metastable WIMP $\eta$, the freeze-out abundance $Y_{\eta \rm FO}$ goes as input here. The left panel of upper row in figure \ref{Fig:compare-Y1} shows the variation of DM relic density as a function of temperature for different values of the freeze-out abundance of $\eta$. From the figure, it is clear that the final abundance of DM is proportional to the freeze-out abundance of $\eta$ as expected. The right panel of upper row in figure \ref{Fig:compare-Y1} shows the variation of DM relic density as a function of temperature for different benchmark values of DM mass by keeping freeze-out abundance of $\eta$ and mass of $\eta$ fixed. As expected, the final DM relic abundance increases for increase in DM mass. In the bottom panel plot of figure \ref{Fig:compare-Y1}, we vary the mother particle's mass while keeping DM mass and mother particle's freeze-out abundance fixed. We see that the final abundance of DM does not depend much upon mother particle's mass as long as the freeze-out abundance remains fixed. From these plots it is clear that the final abundance of DM $\big(\Omega_{DM} h^2 = m_{DM} Y_{DM} s_{0}\big)$ strictly depends on DM mass whereas it is almost independent of the mass of the mother particle. This is due to the fact that $\eta$ has the same branching ratio to two different decay channels (charged lepton pairs and DM plus lepton) therefore half of it's abundance going into that of DM irrespective of its mass. 

\begin{figure}[h]
\includegraphics[width=0.6\textwidth]{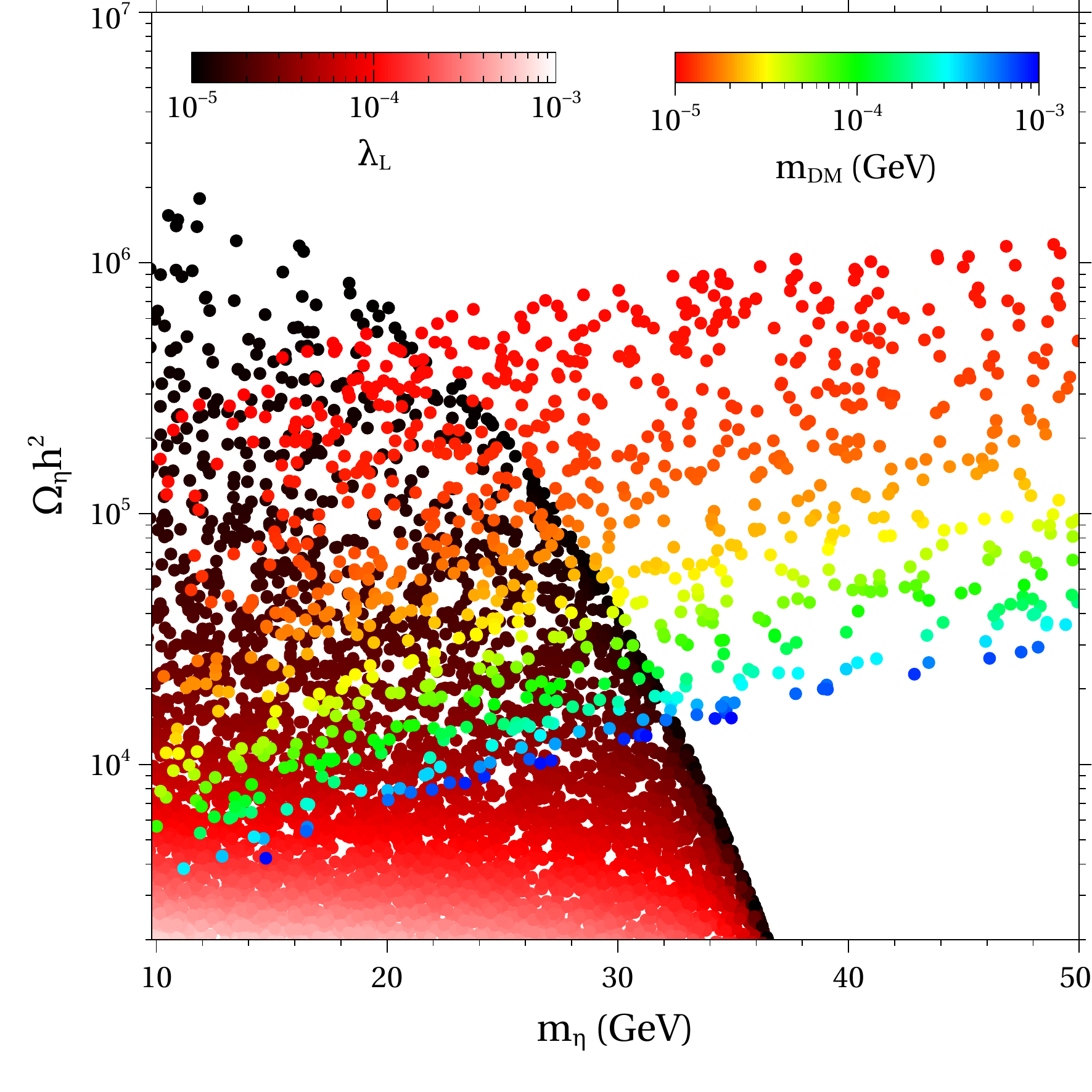}
\caption{Scan of model parameters for case I so that correct DM relic abundance is obtained.}
\label{fig:scancase1}
\end{figure}
We summarise our results for case I as a scan plot shown in figure \ref{fig:scancase1}. The plot shows freeze-out abundance of $\eta$ and mass of $\eta$ from two different directions and requirements. The orange-blue coloured band comes from the relic density requirement of DM. Since DM is being produced from $\eta$ decay after thermal freeze-out of $\eta$, the orange-blue coloured points give us the freeze-out abundance of $\eta$ and its mass so that DM with a particular mass in that coloured band has the correct relic abundance. For this plot, the effective Yukawa coupling between $\eta$ and DM has been kept fixed at $10^{-12}$. As can be seen, the dependence on $\eta$ mass is marginal which was also observed in the benchmark plots shown in figure \ref{Fig:compare-Y1}. This is expected as DM abundance should strongly depend upon the freeze-out abundance of $\eta$ as well as DM mass. 

We then check whether the desired freeze-out abundance of $\eta$ can actually be obtained for some choices of $\eta$ parameters. We can write down the components of $\eta$ as 
\begin{equation}
\eta=\left( \eta^\pm, \;\; \frac{\eta_R+i\eta_I}{\sqrt 2}\right)^T.
\end{equation}
Here we consider 
$\eta_R$ as the lightest component of $\eta$. Calculating the thermal freeze-out abundance of $\eta_R$ is similar to the DM relic calculation in the inert doublet model discussed extensively in2 the literature \cite{Ma:2006km,Barbieri:2006dq, LopezHonorez:2006gr}. Typically there exists two distinct mass regions, $M_{\eta} \leq 80$ GeV and  $M_{\eta} \geq 500$ GeV, where correct relic abundance criteria can be satisfied. In both regions, depending on the mass differences $m_{\eta^\pm}-m_{\eta_R}, m_{\eta_I}-m_{\eta_R}$, the coannihilations of $\eta_R, \eta^\pm$ and $\eta_R, \eta_I$ can also contribute to the DM  relic abundance \cite{Griest:1990kh, Edsjo:1997bg}. The parameters which crucially affect the thermal freeze-out abundance of $\eta$ are the mass splitting $\Delta M_{\eta}$ within the components of $\eta$ and $\eta$ couplings with the SM Higgs $\lambda_L$. It should be noted that there are four different Higgs doublets which can mediate $\eta$ interactions with the SM particles. Here, for simplicity, we consider the scalar interactions to be mediated only by the SM like Higgs, governed by the coupling $\lambda_L$. We fix the mass splitting as $\Delta M_{\eta}=50$ GeV and scan over $\lambda_L$ to find the thermal freeze-out abundance of $\eta_R$. The corresponding region of parameter space is shown as red-black colour coded region of figure \ref{fig:scancase1}. The overlapping region of orange-blue and red-black colour coded regions of figure \ref{fig:scancase1} contain the points which satisfy the correct DM abundance criteria in our model. As can be seen from the colour codes, smaller the DM mass, more overproduced $\eta_R$ has to be, as expected. And as it is well known for freeze-out of scalar doublet, it is the low mass regime of $\eta_R$ where it is more likely to be overproduced due to the absence of sufficient annihilation channels resulting from kinematic suppression as well as departure from s-channel resonance. One can also have such overproduced regime of $\eta_R$ in the high mass limit $ m_{\eta} \geq 1$ TeV which we have not shown in this plot.

It should be noted that, there exists strong bounds on the masses of different components of $\eta$ as well as their couplings with the SM Higgs. For example, precision measurements of $W, Z$ decay widths at LEP experiment lead to the bounds
$$ m_{\eta^\pm}+m_{\eta_R} > M_{W^{\pm}}, \; m_{\eta_I}+m_{\eta_R} > M_Z, \; m_{\eta^\pm}+m_{\eta_I}> M_{W^{\pm}}, 2m_{\eta^\pm} > M_Z $$
The region defined by the intersection of the following conditions
$$ m_{\eta_R} < 80 \; {\rm GeV}, m_{\eta_I} < 100 \; {\rm GeV}, m_{\eta_I}-m_{\eta_R} > 8 \; {\rm GeV}$$
is also excluded from the non-observation of dijet or dilepton signals at LEP \cite{Lundstrom:2008ai, Belyaev:2016lok}. The charged scalars are constrained as $m_{\eta^\pm} > 70$ GeV by reinterpreting the LEP bounds on charginos. We apply these bounds here simply by taking a conservative mass splitting $\Delta M = m_{\eta^\pm}-m_{\eta_R}=m_{\eta_I}-m_{\eta_R} =100$ GeV in figure \ref{fig:scancase1}. Another bound comes from the LHC from the measurements of invisible decay width of the SM Higgs as well as SM Higgs decay into diphoton \cite{Belyaev:2016lok}. We check that the range of parameters we scan through satisfy these bounds.

\subsection{Case II}
\begin{figure}[h!]
		\includegraphics[width=0.45\textwidth]{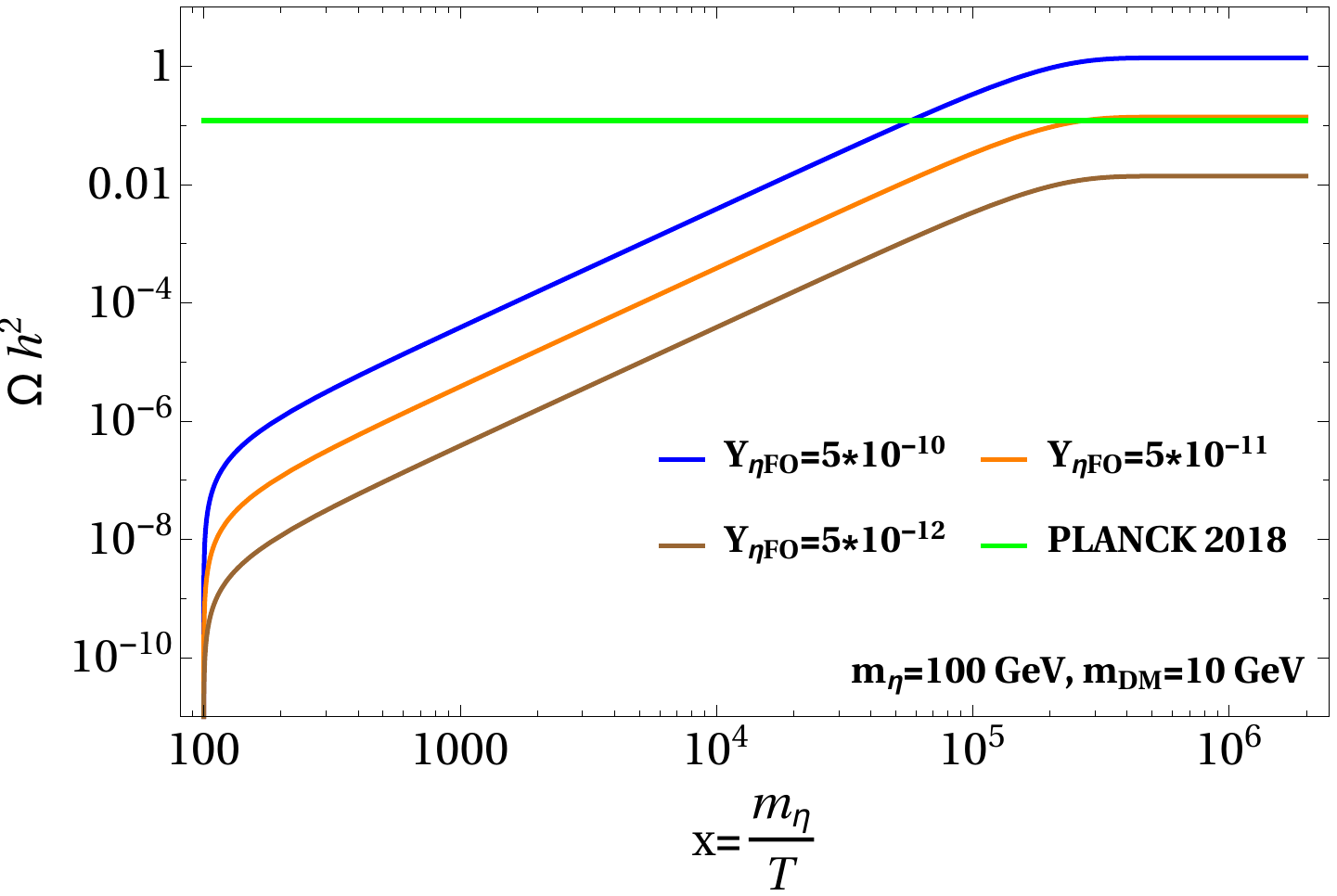}
		\includegraphics[width=0.45\textwidth]{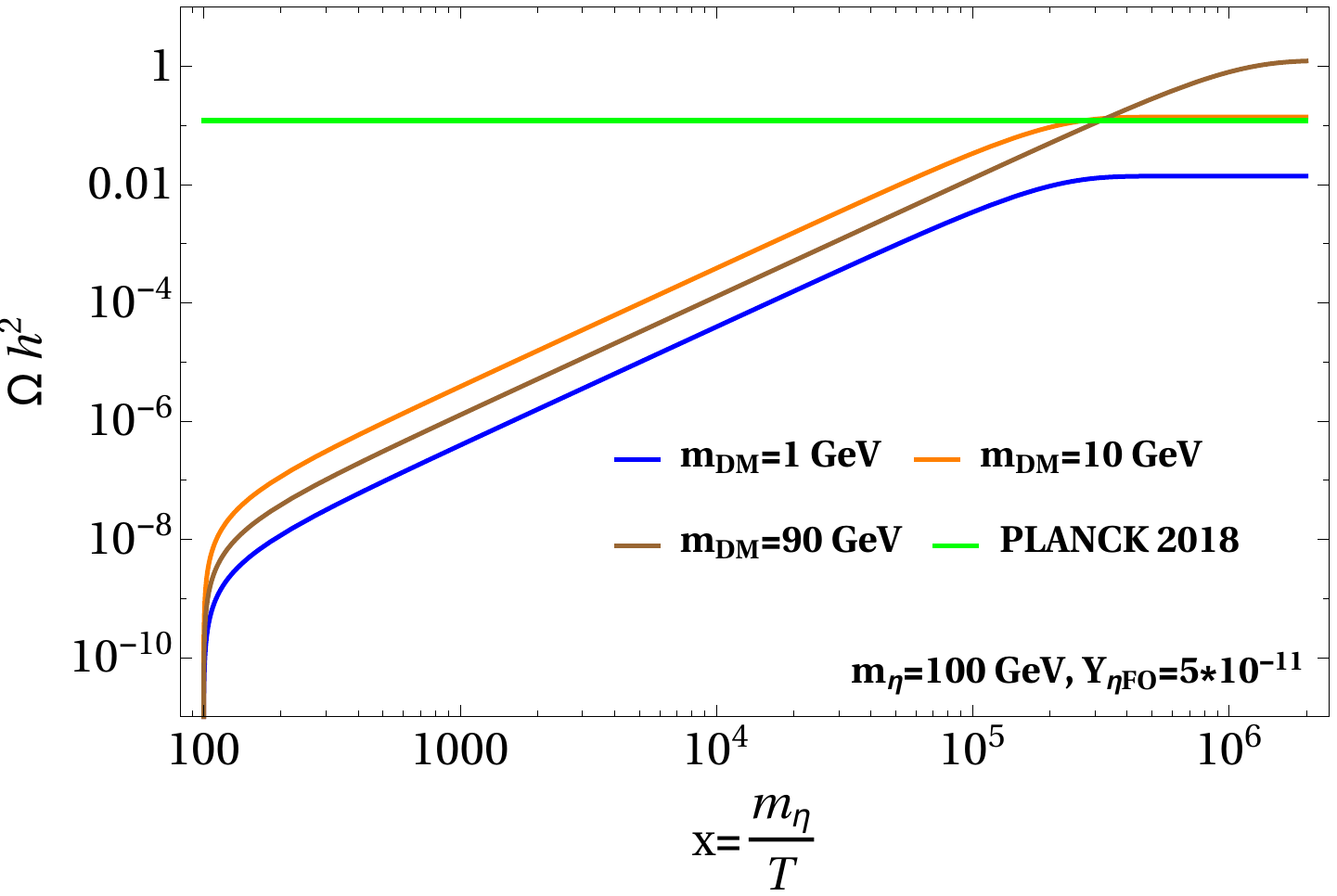} \\
		\includegraphics[width=0.45\textwidth]{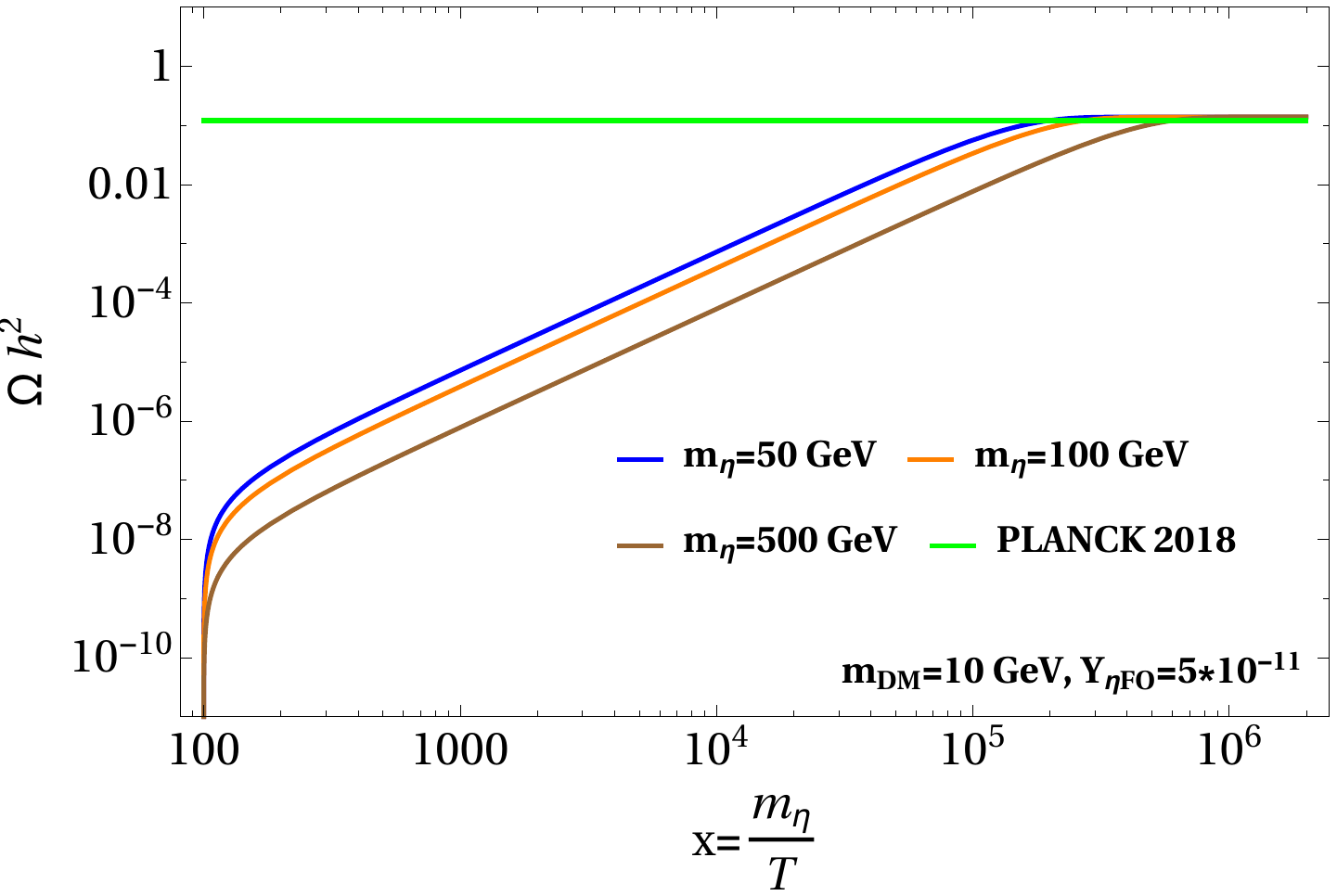}
\caption{Comparison of $\Omega_{DM} h^2$ with respect to different model parameters for case II.}
\label{Fig:compare-Y2}
\end{figure}

In this case $\eta$ can decay only through the interaction $\frac{ \langle \phi_{N_i} \rangle}{M_{Pl}} \eta \bar{\nu} \psi$ as the other interaction is forbidden here. As there is no decay terms present for $\psi$, it is not required to consider sub-MeV mass of $\psi$ for forbid its decay kinematically. We consider DM mass in the GeV regime here. The relevant coupled Boltzmann equations in this case are almost same as the ones discussed in previous subsection namely, \eqref{BEeta:case1} and \eqref{BEpsi:case1} except the fact that $\Gamma_{\eta}$ is now replaced by $\Gamma_{\eta \rightarrow \psi \bar{\nu}}$ as there is no other decay channel present for $\eta$. They can be written as 
\begin{eqnarray}
\frac{dY_{\eta}}{dx} &=&  - \frac{M_{\rm Pl}}{1.66} \frac{x\sqrt{g_{\star}(x)}}{M_{\rm sc}^2\ g_s(x)} \Gamma_{\eta \rightarrow \psi \bar{\nu}}  \ Y_{\eta},
\label{BEeta:case2}
\end{eqnarray}
\begin{eqnarray}
\frac{dY_\psi}{dx} &=& \frac{M_{\rm Pl}}{1.66} \frac{x\sqrt{g_{\star}(x)}}{M_{\rm sc}^2 \ g_s(x)} \Gamma_{\eta \rightarrow \psi \bar{\nu}}
\ Y_{\eta}.
\label{BEpsi:case2}
\end{eqnarray}
In figure \ref{Fig:compare-Y2} we have shown the variation of DM abundance as a function of $(x=\frac{m_{\eta}}{T})$ for different benchmark values of model parameters. The left panel of upper row in figure \ref{Fig:compare-Y2} shows the variation caused by different freeze-out abundance of mother particle while the right panel plot shows the variation with DM mass. In the lower panel plot of figure \ref{Fig:compare-Y2} we show the variation due to change in mother particle's mass. As in case I, the final abundance of DM does not depend much upon mother particle's mass as long as its freeze-out abundance is kept fixed. One interesting feature observed in right panel plot in the upper row of this figure (not noticed in case I) is the change in evolution of DM density as DM mass becomes closer to mother particle's mass. Final abundance of DM is always proportional to DM mass but when $m_{DM}$ becomes very close to the $m_{\eta}$ then $\eta$ decays slowly and that can be seen from the brown line corresponding to $\big( m_{DM}=90 \rm GeV\big)$ which grows slower with $x$ compared to the line corresponding to lower values of DM mass $\big( m_{DM}=10 \rm GeV\big)$.
\begin{figure}[h]
\includegraphics[width=0.6\textwidth]{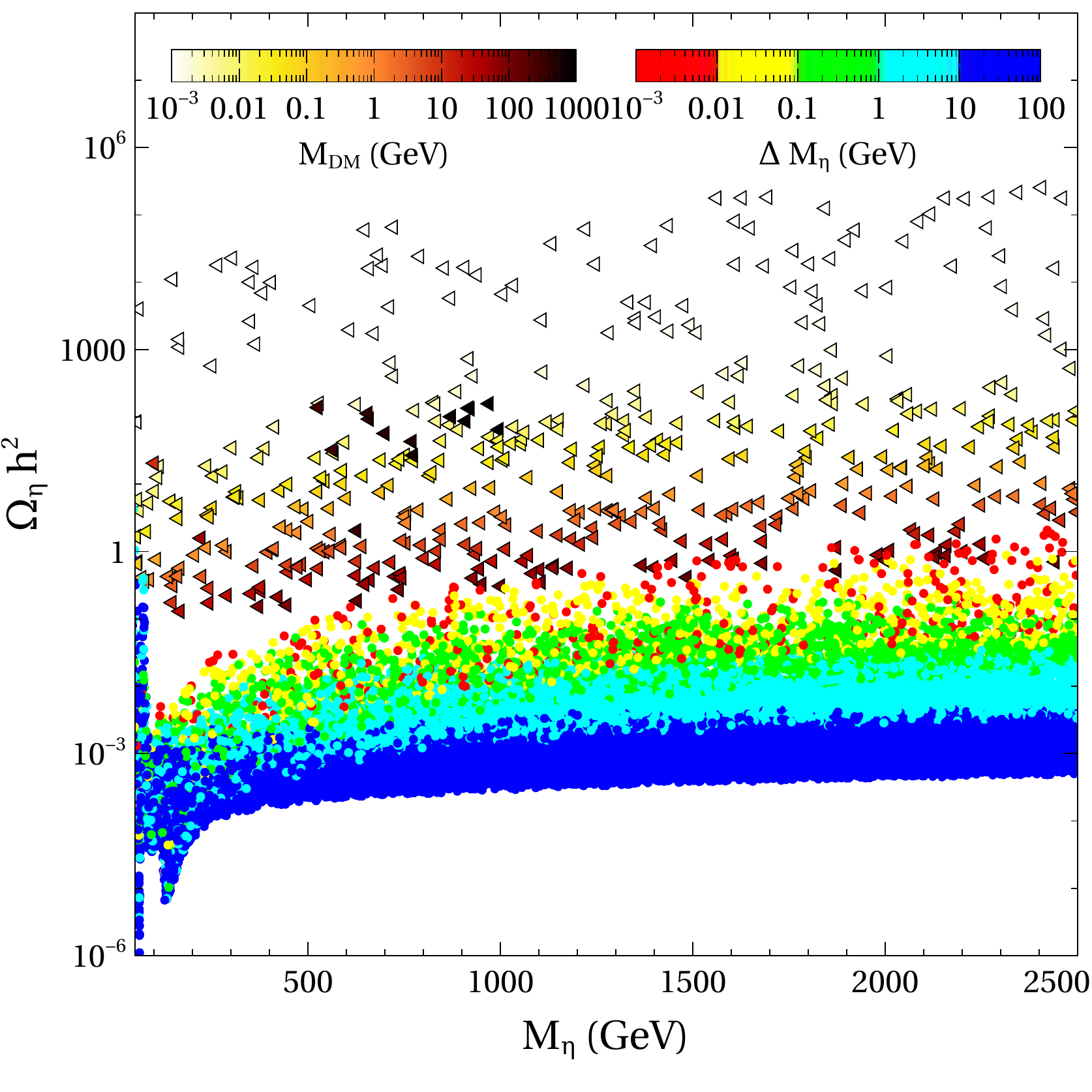}
\caption{Scan of model parameters for case II so that correct DM relic abundance is obtained.}
\label{fig:scancase2}
\end{figure}

Similar to case I, here also we make a parameter scan that can give rise to the correct relic abundance of super-WIMP DM. The plot is shown in figure \ref{fig:scancase2} where the triangles correspond to DM masses in the white-red colour code while dots correspond to $\Delta M_{\eta}$ with blue-red colour code. The triangular points correspond to the parameter space that gives rise to correct DM abundance for a particular set of $(M_{\rm DM}, M_{\eta}, \Omega_{\eta}{\rm h}^2)$. As before, the mass of $\eta$ does not play much role on DM abundance. However, $\Delta M_{\eta}$ plays crucial role in generating the required freeze-out abundance of $\eta$. As can be seen from this plot, there exists very small overlaps between triangles and dots which correspond to correct DM abundance as well as realistic freeze-out abundance of mother particle for particular choice of $\Delta M_{\eta}$. Thus, case II gets more constrained compared to case I in terms of final parameter space. 
\begin{figure}[h!]
		\includegraphics[width=0.45\textwidth]{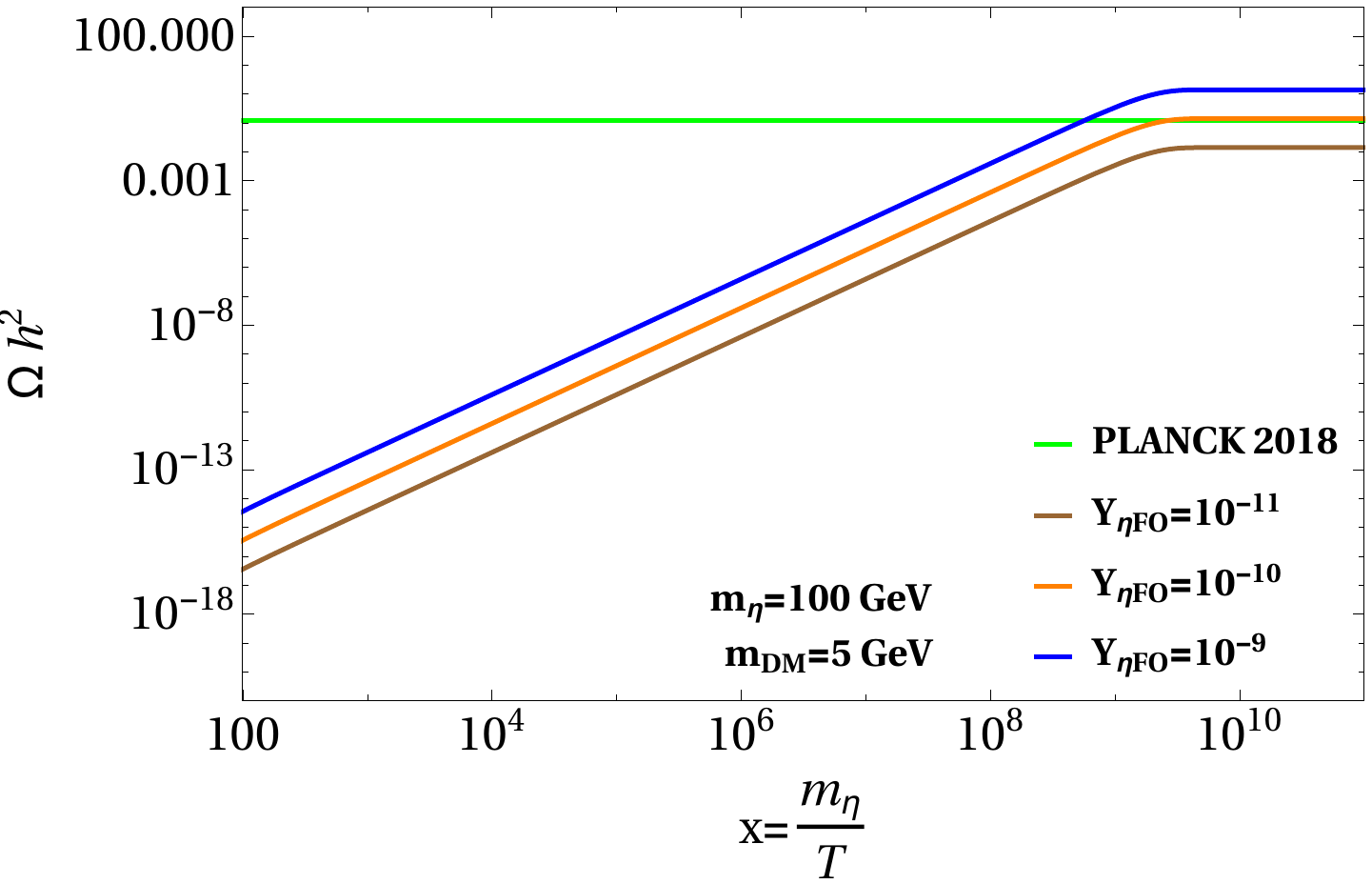}
		\includegraphics[width=0.45\textwidth]{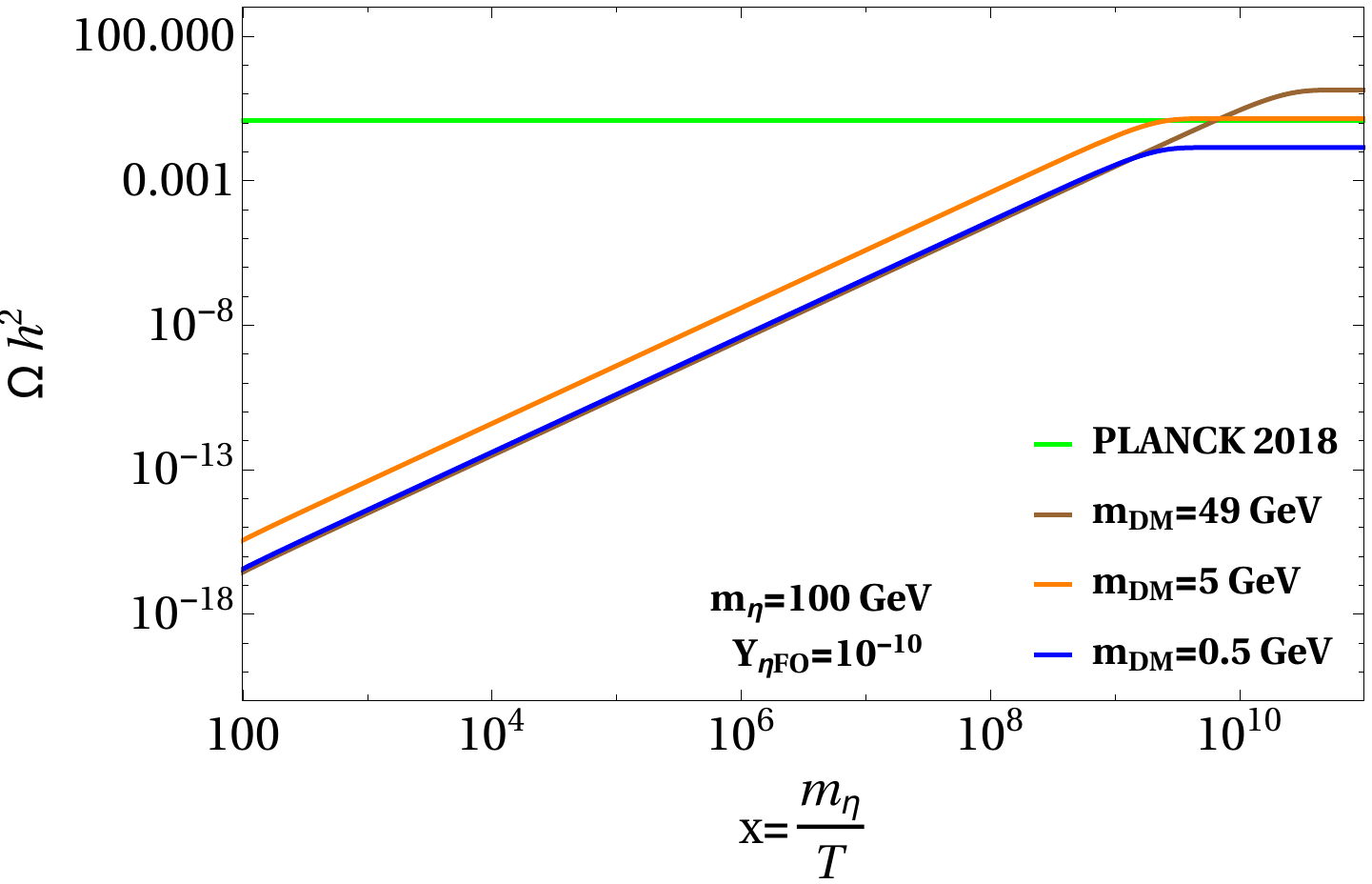} \\
		\includegraphics[width=0.45\textwidth]{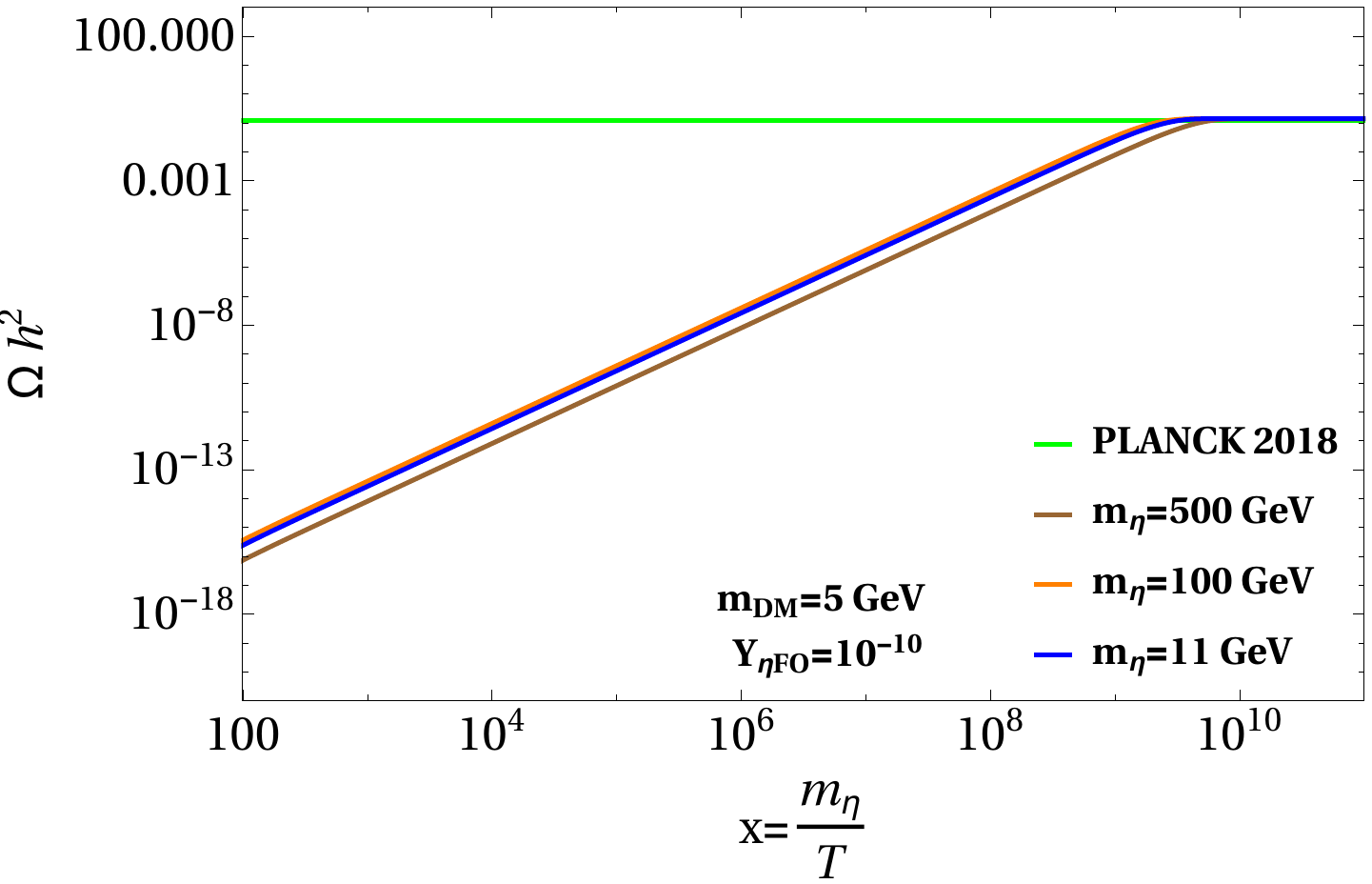}
\caption{Comparison of $\Omega_{DM} h^2$ with respect to different model parameters for case III.}
\label{Fig:compare-Y3}
\end{figure}

\subsection{Case III}
We finally consider the case where the mother particle can decay into DM only through interactions of type $\tilde{Y_1} \frac{ \langle H \rangle}{M_{\rm Pl}} \eta^{\dagger} \psi \psi $. This differs from case II by the facts (i) mother particle decays into two DM particles instead of one, (ii) the effective strength of the vertex $\eta-\psi-\psi$ is much weaker $Y_{\rm eff} \approx \tilde{Y_1} 10^{-17}$. The relevant Boltzmann equations are 
\begin{eqnarray}
\frac{dY_{\eta}}{dx} &=&  - \frac{M_{\rm Pl}}{1.66} \frac{x\sqrt{g_{\star}(x)}}{M_{\rm sc}^2\ g_s(x)} \Gamma_{\eta \rightarrow \psi \bar{\psi}}  \ Y_{\eta},
\label{BEeta:case3}
\end{eqnarray}
\begin{eqnarray}
\frac{dY_\psi}{dx} &=& \frac{2M_{\rm Pl}}{1.66} \frac{x\sqrt{g_{\star}(x)}}{M_{\rm sc}^2 \ g_s(x)} \Gamma_{\eta \rightarrow \psi \bar{\psi}}
\ Y_{\eta}.
\label{BEpsi:case3}
\end{eqnarray}
where the decay width $\Gamma_{\eta \rightarrow \psi \bar{\psi}}$ is given by equation \eqref{decay:psipsi}. The corresponding results are shown in figure \ref{Fig:compare-Y3} where we show the evolution of DM relic with temperature for different benchmark choices of model parameters. The overall evolution remains similar to that in case II (shown in figure \ref{Fig:compare-Y2}) except for the fact that due to smaller effective Yukawa coupling between mother particle and DM, the yield in DM abundance happens at relatively lower temperatures or higher $x$. The parameter space scan for case III remains very similar to that for case II (shown in figure \ref{fig:scancase2}) and we skip it showing again.
\begin{figure}[h]
\includegraphics[width=0.6\textwidth]{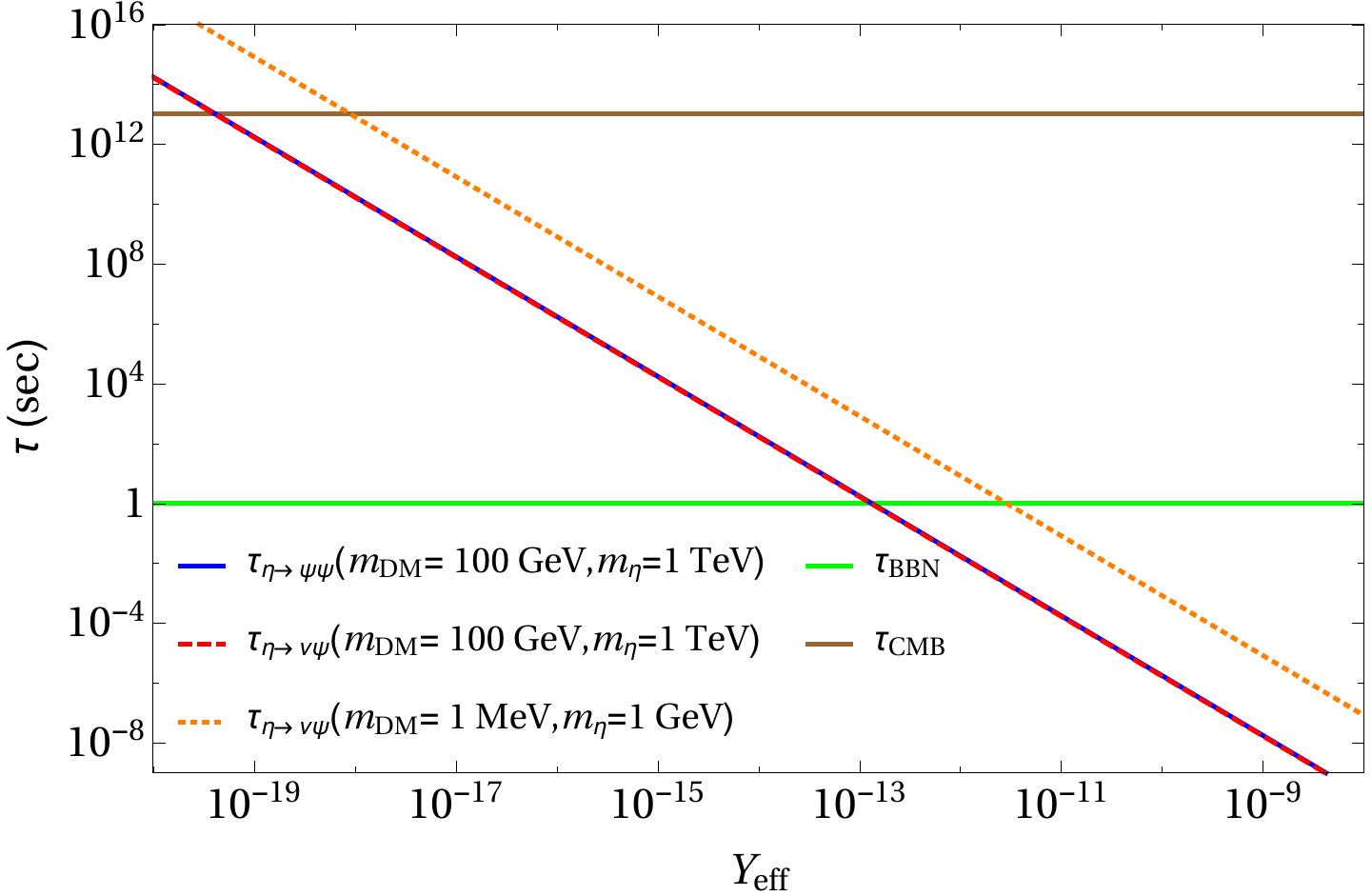}
\caption{Lifetime of mother particle $\eta$ as function of effective Yukawa coupling $Y_{\rm eff}$ for its two body decay into dark matter.}
\label{fig:lifetime}
\end{figure}

It should be noted that in case I and II, we considered the effective Yukawa coupling between $\eta$ and $\psi$ to be $10^{-12}$ which will require some fine tuning in the Yukawa couplings appearing with the Planck suppressed operators if the scale of $A_4$ symmetry that is, $u$ lies close to GUT scale. In case III, the effective Yukawa coupling is very small $10^{-16}$ which arises naturally from the ratio of electroweak scale to Planck scale. In case III, the effective Yukawa coupling of super-WIMP is independent of $A_4$ breaking scale and requires no fine-tuning. The smallness of effective Yukawa couplings chosen here also justifies the validity of super-WIMP formalism where the major non-thermal contribution to DM relic comes after metastable WIMP freezes out. In general non-thermal or FIMP dark matter scenario, contribution to DM can arise while mother particle is in thermal equilibrium as well as after the mother particle freezes out \cite{Hall:2009bx}. Equilibrium contribution dominates for larger Yukawa coupling $Y_{\rm eff} \sim 10^{-9}-10^{-8}$ and for chosen Yukawa couplings here, the equilibrium contribution remains suppressed below $5\%$. This also has helped us to solve the coupled Boltzmann equations at two stages: first solving only the mother particle's equation for its thermal freeze-out and then solving the coupled Boltzmann equations for mother particle and DM as discussed above. 

Finally, we check whether the three scenarios discussed in our work could affect big bang nucleosynthesis (BBN) or cosmic microwave background (CMB), the two pillars of success of standard $\Lambda$CDM cosmology. The bounds can, in principle, be severe in case I where there is a late decay of mother particle $\eta$ into standard model leptons and dark matter. For example, if $\eta$ decays into light neutrinos and dark matter around the BBN epoch ($t_{\rm BBN} \sim 1 {\rm s}$), it may affect the neutrino decoupling temperature, thereby affecting the successful predictions of BBN. Similarly, if lifetime of $\eta$ is more than the epoch of recombination ($t_{\rm CMB} \sim 10^{13} {\rm s}$), the CMB spectrum can be distorted. Since CMB power spectrum is sensitive to both dark and visible matter, both case I (visible decay) as well as case II, III (invisible decay) can affect it. We plot lifetime of $\eta$ with effective Yukawa coupling $Y_{\rm eff}$ involved in visible decay $\eta \rightarrow {\rm v} \psi, {\rm v} \equiv {\rm SM \; lepton}$ as well as invisible decay $\eta \rightarrow \psi \psi$ and show it in figure \ref{fig:lifetime}. As can be seen from this plot, the effective Yukawa coupling of our case I $Y_{\rm eff} = 10^{-12}$ gives rise to lifetime less than the BBN epoch keeping it safe from relevant constraints. Similarly, the Yukawa coupling of case III that is, $Y_{\rm eff} = 10^{-16}$ gives rise to lifetime less than the recombination epoch, evading tight constraints on mass difference $\delta m = m_{\eta}-m_{\rm DM}$ for both visible decay \cite{DEramo:2018khz} as well as invisible decay \cite{Enqvist:2015ara, Cheng:2015dga}.

\section{Conclusion}
\label{sec4}
We have studied a scenario where light neutrino masses and mixing, at renormalisable level, is dictated by  discrete flavour symmetries based on $A_4 \times Z_4$. While the non-Abelian discrete flavour symmetry $A_4$ leads to TBM type light neutrino mixing for generic choices of vacuum alignments, the $Z_4$ symmetry dictates the dynamics of dark sector comprising of a scalar doublet $\eta$ and a singlet vector like fermion $\psi$. Clearly the renormalisable version of the model is ruled out by the observations of non-zero reactor mixing angle $\theta_{13}$. Also the dark sector it predicts can have a inert scalar doublet dark matter with correct relic abundance while the singlet fermion in the dark sector remains decoupled from the usual thermal bath resulting in its negligible abundance.

In order to achieve the correct neutrino phenomenology, we utilise the fact that global symmetries are conjectured to be broken explicitly at Planck scale, possibly by gravitational effects. Such effects can mimic as Planck suppressed operators in the model which explicitly break the discrete symmetries. These corrections not only can generate non-zero $\theta_{13}$ but also changes the dark sector dynamics. In the minimal scenario, such Planck suppressed operators can induce decay of dark sector scalar doublet $\eta$ into both standard model particles as well as the dark sector singlet fermion $\psi$. In fact such operators also open up decay modes of dark sector particles into the SM ones. In the minimal model, one can make $\psi$ kinematically long lived by going to low mass range and tuning the relevant couplings. On the other hand, the decay modes of $\eta$ into $\psi$ helps in realising the super-WIMP mechanism where a metastable WIMP freezes-out from the thermal bath and then decays into a super-weakly interacting dark matter particle.

The tree level neutrino mass matrix originating from a type I seesaw mechanism can receive several corrections due to the Planck
suppressed operators. Such corrections can arise either in the light neutrino mass matrix directly due to Weinberg operator, 
in Dirac neutrino mass matrix or heavy right handed neutrino mass matrix out of which the first one is negligible compared to
the latter ones. To illustrate the role of such corrections in a simple manner, we only take the corrections to the Dirac 
neutrino mass matrix and show that the corrections from Planck suppressed operators can generate the necessary deviations from
TBM mixing leading to a non-zero value of $\theta_{13}$, in agreement with observations. Owing to the specific flavor structure of the model we have specific correlation among
the mixing angles appearing in the lepton mixing matrix. Such correlations can be tested in
future neutrino oscillation experiments like DUNE, T2HK etc\cite{Agarwalla:2017wct, Petcov:2018snn}. However
a detailed study in this direction is beyond the scope of present study. We also outline the super-WIMP dark matter phenomenology by considering three distinct scenario: (i) $\eta$ decays to SM particles as well as $\psi$ and $\psi$ is kinematically long lived, (ii) $\eta$ decays to $\psi$ and a SM neutrino while $\psi$ is perfectly stable, (iii) $\eta$ decays into a pair of $\psi$ while $\psi$ is perfectly stable. Out of these, the first scenario correspond to the model which we have discussed in our work while the latter two scenario can be realised if the discrete $Z_4$ symmetry in the dark sector can be uplifted to a gauge symmetry which does not get broken by gravity effects. While we do not discuss such UV complete gauge symmetric realisation of $Z_4$ symmetry we outline the interesting differences for super-WIMP phenomenology. The analysis for neutrino sector in all three DM scenarios remain same, however. 

We show that correct dark matter relic abundance can be obtained in all three distinct scenarios. While direct detection of super-WIMP dark matter itself may not be very optimistic, the mother particle $\eta$ can be probed at ongoing experiments. Since the lightest component of $\eta$ decays to DM as well as SM leptons (in case I) with very feeble couplings one can probe them in colliders either as missing energy or displaced vertex, long charged tracks depending upon the lifetime. If charged component of $\eta$ is the lightest component of $\eta$ then for typical super-WIMP couplings discussed in this work, it will give rise to a long charged track in colliders as its decay length will be much larger than typical displaced vertex ones searched for. For more discussions on displaced vertex and disappearing charged track signatures of a similar model, please refer to a recent work \cite{Borah:2018smz}. Apart from signatures of mother particle, the DM itself can leave some detectable signatures, specially in case I where it is not perfectly stable but long lived. For example, a 7 keV long-lived fermion DM\footnote{More details of such keV fermion dark matter can be found in a recent review \cite{Adhikari:2016bei}.} can decay into a photon and light neutrino at radiative level with $W$ boson and charged leptons of the SM in loop. The corresponding decay width is given by \cite{Pal:1981rm}
\begin{equation}
\Gamma (\psi \rightarrow \nu \gamma) \approx 1.38 \times 10^{-29} \; \text{s}^{-1} \left ( \frac{\sin^2{2\theta}}{1\times 10^{-7}} \right) \left ( \frac{M_{\psi}}{1 \; \text{keV}} \right)^5
\end{equation}
where $\theta$ denotes the mixing between $N_1$ and $\nu$. From the observation of the 3.55 keV line, which can arise from the decay of a 7.1 keV sterile neutrino DM, the mixing angle which is in agreement with the observed flux is $\sin^2{2 \theta} \approx 7 \times 10^{-11}$ \cite{Bulbul:2014sua}. Such a mixing angle can be naturally obtained in the model, as seen from the expression for mixing angle given in equation \eqref{mixing1}. Although the analysis of the preliminary data collected by the Hitomi satellite (before its unfortunate crash) do not confirm such a monochromatic line \cite{Aharonian:2016gzq}, one still needs to wait for a more sensitive observation with future experiments to have a final word on it. We leave a more detailed study of detection prospects as well as UV complete realisations of case II, III to future works.

\acknowledgments
DB acknowledges the support from IIT Guwahati start-up grant (reference number: xPHYSUGI-ITG01152xxDB001), Early Career Research Award from DST-SERB, Government of India (reference number: ECR/2017/001873) and Associateship Programme of Inter University Centre for Astronomy and Astrophysics (IUCAA), Pune. DB is also grateful to the Mainz Institute for Theoretical Physics (MITP) of the DFG Cluster of Excellence ${\rm PRISMA}^+$ (Project ID 39083149), for its hospitality and its partial support during the completion of this work. DN would like to thank Anirban Biswas for useful discussions.

\appendix
\section{$A_4$ Multiplication Rules}
\label{appen1}
The non-Abelian discrete group $A_4$, also the symmetry group of a tetrahedron, is a group of 
even permutations of four objects. This group has four irreducible representations out of which 
three are one-dimensional and one three dimensional, denoted by $\bf{1}, 
\bf{1'}, \bf{1''}$ and $\bf{3}$ respectively, being consistent with the sum of 
square of the dimensions $\sum_i n_i^2=12$. We denote a generic permutation 
$(1,2,3,4) \rightarrow (n_1, n_2, n_3, n_4)$ simply by $(n_1 n_2 n_3 n_4)$. The 
group $A_4$ can be generated by two basic permutations $S$ and $T$ given by $S = 
(4321), T=(2314)$. This satisfies 
$$ S^2=T^3 =(ST)^3=1$$
which is called a presentation of the group. Their product rules of the 
irreducible representations are given as
$$ \bf{1} \otimes \bf{1} = \bf{1}$$
$$ \bf{1'}\otimes \bf{1'} = \bf{1''}$$
$$ \bf{1'} \otimes \bf{1''} = \bf{1} $$
$$ \bf{1''} \otimes \bf{1''} = \bf{1'}$$
$$ \bf{3} \otimes \bf{3} = \bf{1} \otimes \bf{1'} \otimes \bf{1''} \otimes 
\bf{3}_a \otimes \bf{3}_s $$
where $a$ and $s$ in the subscript corresponds to anti-symmetric and symmetric 
parts respectively. Denoting two triplets as $(a_1, b_1, c_1)$ and $(a_2, b_2, 
c_2)$ respectively, their direct product can be decomposed into the direct sum 
mentioned above. In the $S$ diagonal basis, the products are given as
$$ \bf{1} \backsim a_1a_2+b_1b_2+c_1c_2$$
$$ \bf{1'} \backsim a_1 a_2 + \omega^2 b_1 b_2 + \omega c_1 c_2$$
$$ \bf{1''} \backsim a_1 a_2 + \omega b_1 b_2 + \omega^2 c_1 c_2$$
$$\bf{3}_s \backsim (b_1c_2+c_1b_2, c_1a_2+a_1c_2, a_1b_2+b_1a_2)$$
$$ \bf{3}_a \backsim (b_1c_2-c_1b_2, c_1a_2-a_1c_2, a_1b_2-b_1a_2)$$
In the $T$ diagonal basis on the other hand, they can be written as
$$ \bf{1} \backsim a_1a_2+b_1c_2+c_1b_2$$
$$ \bf{1'} \backsim c_1c_2+a_1b_2+b_1a_2$$
$$ \bf{1''} \backsim b_1b_2+c_1a_2+a_1c_2$$
$$\bf{3}_s \backsim \frac{1}{3}(2a_1a_2-b_1c_2-c_1b_2, 2c_1c_2-a_1b_2-b_1a_2, 
2b_1b_2-a_1c_2-c_1a_2)$$
$$ \bf{3}_a \backsim \frac{1}{2}(b_1c_2-c_1b_2, a_1b_2-b_1a_2, c_1a_2-a_1c_2)$$

Denoting two triplets as $(a_1, b_1, c_1)$ and $(a_2, b_2, c_2)$ respectively, their direct product in the T diagonal basis can be decomposed into the direct sum as
$$ \bf{1} \backsim a_1a_2+b_1c_2+c_1b_2$$
$$ \bf{1'} \backsim c_1c_2+a_1b_2+b_1a_2$$
$$ \bf{1''} \backsim b_1b_2+c_1a_2+a_1c_2$$
$$\bf{3}_s \backsim (2a_1a_2-b_1c_2-c_1b_2, 2c_1c_2-a_1b_2-b_1a_2, 2b_1b_2-a_1c_2-c_1a_2)$$
$$ \bf{3}_a \backsim (b_1c_2-c_1b_2, a_1b_2-b_1a_2, c_1a_2-a_1c_2)$$


\end{document}